\documentclass[%
floatfix,
showkeys,
twocolumn, %
nofootinbib, %
superscriptaddress, %
longbibliography, %
]{revtex4-1}

\usepackage{cmap}

\usepackage{ucs}
\usepackage[utf8x]{inputenc}
\usepackage[T1,T2A]{fontenc}
\usepackage[german,russian,english]{babel}

\usepackage[sort&compress]{natbib}
\usepackage{amsmath}
\usepackage{amssymb}

\usepackage{mathtools}
\mathtoolsset{
showonlyrefs,
mathic = true
}

\allowdisplaybreaks

\usepackage{hyperref}
\hypersetup{backref,
 colorlinks=false}
\hypersetup{pdfborder=0 0 0}

\usepackage{breakurl}

\usepackage{microtype}
\UseMicrotypeSet[protrusion]{alltext}

\usepackage{listings}
\usepackage{listingsutf8}
\usepackage{nicefrac}

\usepackage{comment}
\usepackage{physics}
\usepackage{tensor}

\begin{document}
\graphicspath{{image/weibull-test/en/}}

\title{Approaches to Stochastic Modeling of Wind Turbines}

\author{M. N. Gevorkyan}
\email{gevorkyan_mn@rudn.university}
\affiliation{Department of Applied Probability and Informatics,\\
  Peoples' Friendship University of Russia (RUDN University),\\
  6 Miklukho-Maklaya str., Moscow, 117198, Russia}

\author{A. V. Demidova}
\email{demidova_av@rudn.university}
\affiliation{Department of Applied Probability and Informatics,\\
  Peoples' Friendship University of Russia (RUDN University),\\
  6 Miklukho-Maklaya str., Moscow, 117198, Russia}

\author{Robert A. Sobolewski}
\email{r.sobolewski@pb.edu.pl}
\affiliation{Department of Power Engineering, Fotonics and Lighting Technology,\\
  Bialystok University of Technology,\\
  Poland, 15-351 Bialystok, Wiejska 45 D street}

\author{I. S. Zaryadov}
\email{zaryadov_is@rudn.university}
\affiliation{Department of Applied Probability and Informatics,\\
  Peoples' Friendship University of Russia (RUDN University),\\
  6 Miklukho-Maklaya str., Moscow, 117198, Russia}
\affiliation{Institute of Informatics Problems,
  FRC CSC RAS, IPI FRC CSC RAS,\\
  44-2 Vavilova str.,  Moscow, 119333, Russia}

\author{A. V. Korolkova}
\email{korolkova_av@rudn.university}
\affiliation{Department of Applied Probability and Informatics,\\
  Peoples' Friendship University of Russia (RUDN University),\\
  6 Miklukho-Maklaya str., Moscow, 117198, Russia}

\author{D. S. Kulyabov}
\email{kulyabov_ds@rudn.university}
\affiliation{Department of Applied Probability and Informatics,\\
  Peoples' Friendship University of Russia (RUDN University),\\
  6 Miklukho-Maklaya str., Moscow, 117198, Russia}
\affiliation{Laboratory of Information Technologies\\
  Joint Institute for Nuclear Research\\
  6 Joliot-Curie, Dubna, Moscow region, 141980, Russia}

\author{L. A. Sevastianov}
\email{sevastianov_la@rudn.university}
\affiliation{Department of Applied Probability and Informatics,\\
  Peoples' Friendship University of Russia (RUDN University),\\
  6 Miklukho-Maklaya str., Moscow, 117198, Russia}
\affiliation{Bogoliubov Laboratory of Theoretical Physics\\
  Joint Institute for Nuclear Research\\
  6 Joliot-Curie, Dubna, Moscow region, 141980, Russia}

\begin{abstract}

\textbf{Background.} This paper study statistical data gathered from wind turbines located on 
the territory of the Republic of Poland. The research is aimed to construct the stochastic model that 
predicts the change of wind speed with time.
\textbf{Purpose.} The purpose of this work is to find the optimal distribution for the 
approximation of available statistical data on wind speed.
\textbf{Methods.} We consider four distributions of a random variable: Log-Normal, Weibull, 
Gamma and Beta. In order to evaluate the parameters of distributions  we use method of maximum likelihood. 
To assess the the results of approximation we use a quantile-quantile plot.
\textbf{Results.} All the considered distributions properly approximate the available data. The 
Weibull distribution shows the best results for the extreme values of the wind speed.
\textbf{Conclusions.} The results of the analysis are consistent with the common practice 
of using the Weibull distribution for wind speed modeling. In the future we plan to compare 
the results obtained with a much larger data set as well as to build a stochastic model 
of the evolution of the wind speed depending on time.

\end{abstract}


\maketitle

\section{Introduction}

This work is devoted to the problem of stochastic modeling of speed of wind,
 which is used to generate electrical power in wind plants located 
 on the territory of the Republic of Poland. As a first step several 
 distributions for accuracy of the wind speed approximation will be examined. For this purpose 
 Log-Normal, Weibull, Gamma and Beta are chosen. All these
distributions have shape-location-scale parametrisation. For statistical data processing
the authors used Python 3 with
\texttt{numpy}, \texttt{scipy.stats}~\cite{L_scipy} and
\texttt{matplotlib}~\cite{L_matplotlib} libraries and also 
\texttt{Jupyter}~\cite{L_jupyter} --- an interactive shell.

We used books~\cite{L_Johnson1_en,L_Johnson2_en,L_Nelson} as reference 
materials for distributions properties. 
Articles~\cite{L_Frechet,L_Weibull} are the primary sources in which the 
Weibull distribution is presented for the first time. 
Articles~\cite{L_Lun2000145,L_Seguro200075,L_1983WiEng,L_4488041,L_Islam 
2011985,L_Garcia1998139} describe the use of the Weibull distribution in 
the modeling of wind turbines and wind speed.

\label{sec:intro}

\section{The description of the statistical data structure}

The set of statistical data  is stored in the file \texttt{csv} consisting of the following columns:
\begin{enumerate}
  \item $T$ --- time of fixation of wind speed and direction  by sensors installed on the wind power turbine (hh:mm format);
  \item $X_1$ --- output power of wind turbine [kW] (the negative values mean the power is consumed rather then generated);
  \item $X_2$ --- wind speed [m/s] (measured by anemometer installed at the top of wind turbine nacelle);
  \item $X_3$ --- wind direction [deg] (measured by anemometer installed at the top of wind turbine nacelle; measured clockwise, the value 0 to the N);
  \item $X_4$ --- wind speed 10 m [m/s] obtained at 10 m above the ground  m;
  \item $X_5$ --- wind direction 10 m [deg] (obtained at 10 m above the ground; measured clockwise, the value from 0 to the N);
  \item $X_6$ --- wind speed 50 m [m/s] (obtained at 50 m above the ground);
  \item $X_7$ --- wind direction 50 m [deg] (obtained at 50 m above the ground; measured clockwise, the value ftom 0 to the N).
\end{enumerate}
The indicators of wind speed and direction  were read out from the sensors every 10 minutes for about 9 months. 
In total, the table contains 39606 entries.

To make an initial choice of distributions that may be suitable for wind speed approximation, the histograms
of wind speed are drawn. Visual assessment of these histograms suggest that the adequate choice  
will be a "heavy-tailed" distribution. But for the wind direction approximation these distributions are not suitable, 
as can be seen from the figure~\ref{fig: winddir}.

\begin{figure}
  \centering
  \includegraphics[width=0.9\linewidth]{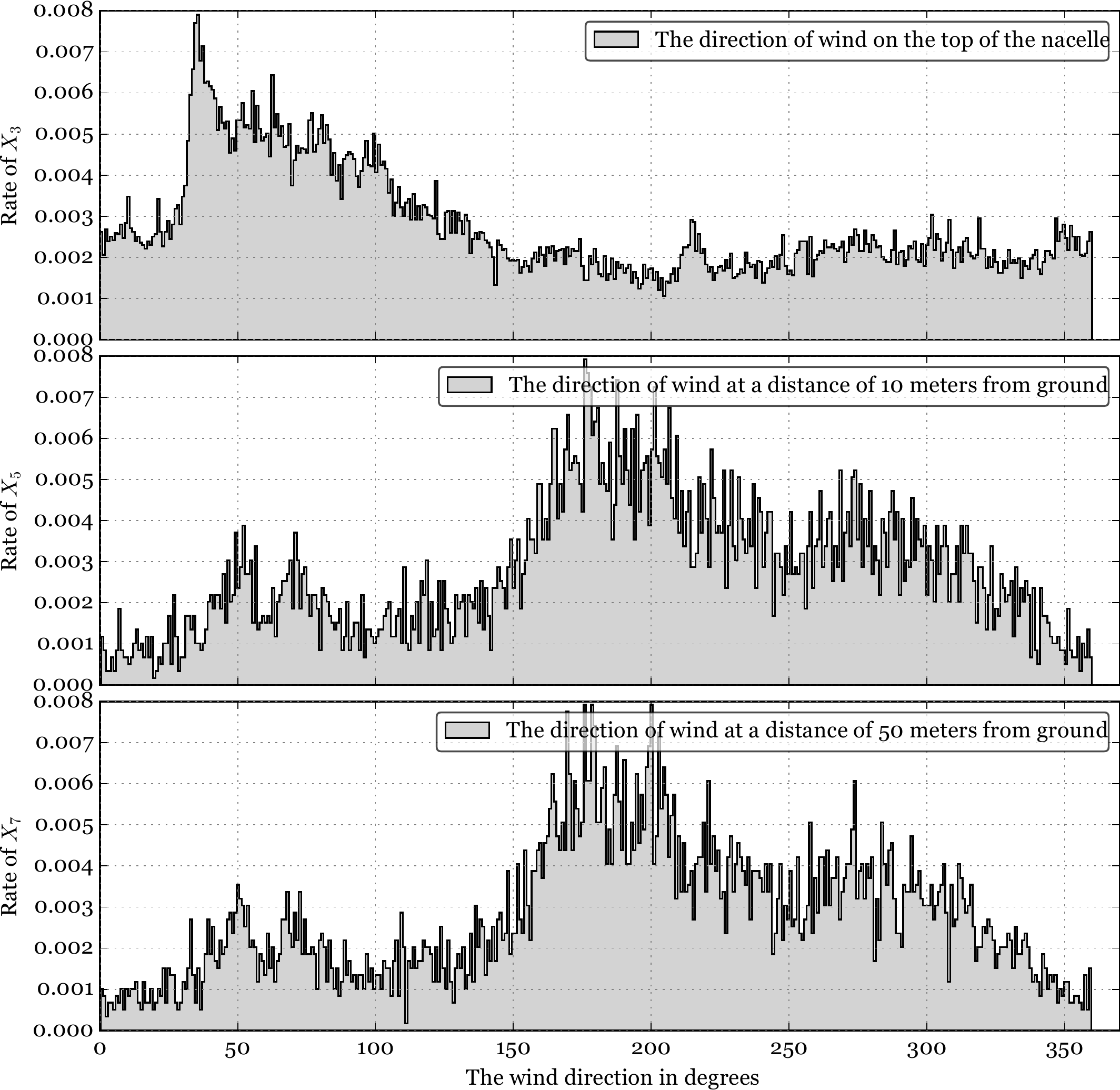}
  \caption{Histogram of wind direction at three levels of height}
  \label{fig:winddir}
\end{figure}

To read out the data we used the function \texttt{genfromtxt} from \texttt{numpy}~\cite{L_scipy} lib.
\lstset{language=Python,breaklines=true,breakatwhitespace=true}
\begin{lstlisting}
ws1, ws2, ws3 = np.genfromtxt('data.csv', delimiter=';', skip_header=True, usecols=(2, 4, 6), unpack=True)
\end{lstlisting}
where \texttt{'data.csv'} is data file, \texttt{delimiter=';'} is
columns separator, \texttt{skip\_header = True} specifies ignoring of
the first line as the names of the columns, \texttt{usecols=(2, 4, 6)}
makes function to use only 2, 4, 6 columns (numbering begins with
zero) and \texttt{unpack=True} --- contents of each column
should be written in separate arrays \texttt{ws1}, \texttt{ws2} and
\texttt{ws3} for further analysis of the data separately.

\section{Probability distributions}

Each of distributions is parameterized by three parameters: $\alpha$
--- shape factor, $l$ --- location factor and $s$ --- scale factor. In
the case of the beta distribution the second scale factor is added,
denoted by $\beta$-letter. All distributions parameters are positive
real numbers: $\alpha,\beta,s,l \in \mathbb{R}$, $\alpha, \beta, s >
0$, $l\geqslant 0$.

The probability density function (PDF) of a Log-Normal random variable $X$ is:
\begin{multline}
  f_{LN}(x; \alpha, l, s)
  = {} \\ {} =
  \begin{dcases}
    \frac{1}{(x-l)\alpha \sqrt{2\pi}}\exp\left(-\frac{1}{2} \left(
        \frac{\ln(x-l) - \ln{s}}{\alpha} \right)^2 \right), & x
    \geqslant l. \\
    0, & x < l.
  \end{dcases}
\end{multline}

The probability density function of a Weibull~\cite{L_Frechet,L_Weibull} random variable $X$ is:
\begin{multline}
  f_{W}(x; \alpha, l, s)
  = {} \\ {} =
  \begin{dcases}
    \frac{\alpha}{s}\left(\frac{x-l}{s}\right)^{\alpha - 1} \exp\left[-\left(\frac{x-l}{s}\right)^{\alpha}\right], & x \geqslant l.\\
    0, & x < l.
  \end{dcases}
\end{multline}

The probability density function of a Gamma random variable $X$ is:
\begin{equation}
  f_{\Gamma}(x; \alpha, l, s) = 
  \begin{dcases}
    \frac{(x-l)^{\alpha-1}\exp\left(-\frac{(x-l)}{s}\right)}{s^{\alpha}\Gamma(\alpha)}, & x \geqslant l.\\
    0, & x < l.
  \end{dcases}
\end{equation}
where $\Gamma(\alpha)$ is gamma-function.

The probability density function of a Beta random variable $X$ is:
\begin{multline}
  f_{\mathcal{B}}(x; \alpha, \beta, l, s)
  = {} \\ {} =
  \begin{dcases}
    \frac{\Gamma(\alpha + \beta)}{s\Gamma(\alpha)\Gamma(\beta)}\left(\dfrac{x-l}{s}\right)^{\alpha - 1}\left(1 - 
        \frac{x - l}{s}\right)^{\beta - 1}, & x \geqslant l.\\
    0, & x < l.
  \end{dcases}
\end{multline}

If in PDF formulas of Log-Normal, Weibull and Gamma distributions let
$l=0$, and for Beta distribution let $s=1$, we get the formulas of
distributions most frequently used in~\cite{L_Johnson1_en,L_Nelson}.

\section{Determination of distributions parameters}

In \texttt{scipy.stats}~\cite{L_scipy} following objects are defined:
\texttt{lognorm}, \texttt{weibull\_min}, \texttt{gamma} and
\texttt{beta}. These objects implement distributions we work
with. Every one of these objects has PDF function \texttt{pdf(x, a,
  [b,] loc, scale)} and CDF (cumulative distribution) function \texttt{cdf(x, a, [b,] loc,
  scale)}, where \texttt{x} --- function argument, \texttt{a},
\texttt{b} --- shape parameters $\alpha$, (and $\beta$ for
Beta-distribution), \texttt{loc} and \texttt{scale} are location and
scale parameters.

For parameters estimation of our distributions the library \texttt{scipy.stats}
provides the function \texttt{fit(data)}, which calculates  the 
parameters of distributions by maximum likelihood method and the empirical data. We used this function to calculate parameters of
the considered distributions. Then we used \texttt{pdf} and
\texttt{cdf} functions to compute values of the probability density
function and cumulative distribution function.

There is the example of the code for the case of Log-Normal distribution:
\begin{lstlisting}
s, loc, scale = scipy.stats.lognorm.fit(ws1)
xs = np.linspace(np.min(ws1), np.max(ws1), 1000)
logN_PDF = scipy.stats.lognorm.pdf(xs, s, loc, scale)
logN_CDF = scipy.stats.lognorm.cdf(xs, s, loc, scale)
\end{lstlisting}
    % \hat{\mu} = \dfrac{1}{n} \sum\limits^{N}_{k=1}\ln X_{k},\;\; \hat{\sigma}^{2} = \dfrac{1}{n}\sum^{N}_{k=1}(\ln X_{k} - \hat{\mu})^2

  % \hat{k}^{-1} = \dfrac{\sum\limits_{i=1}^n x_{i}^{\hat{k}} \ln{x_i}}{\sum\limits_{i=1}^n x_{i}^{\hat{k}}} - \dfrac{1}{n} \sum\limits_{i=1}^n \ln{x_i}

\begin{figure}
  \centering
  \includegraphics[width=0.8\linewidth]{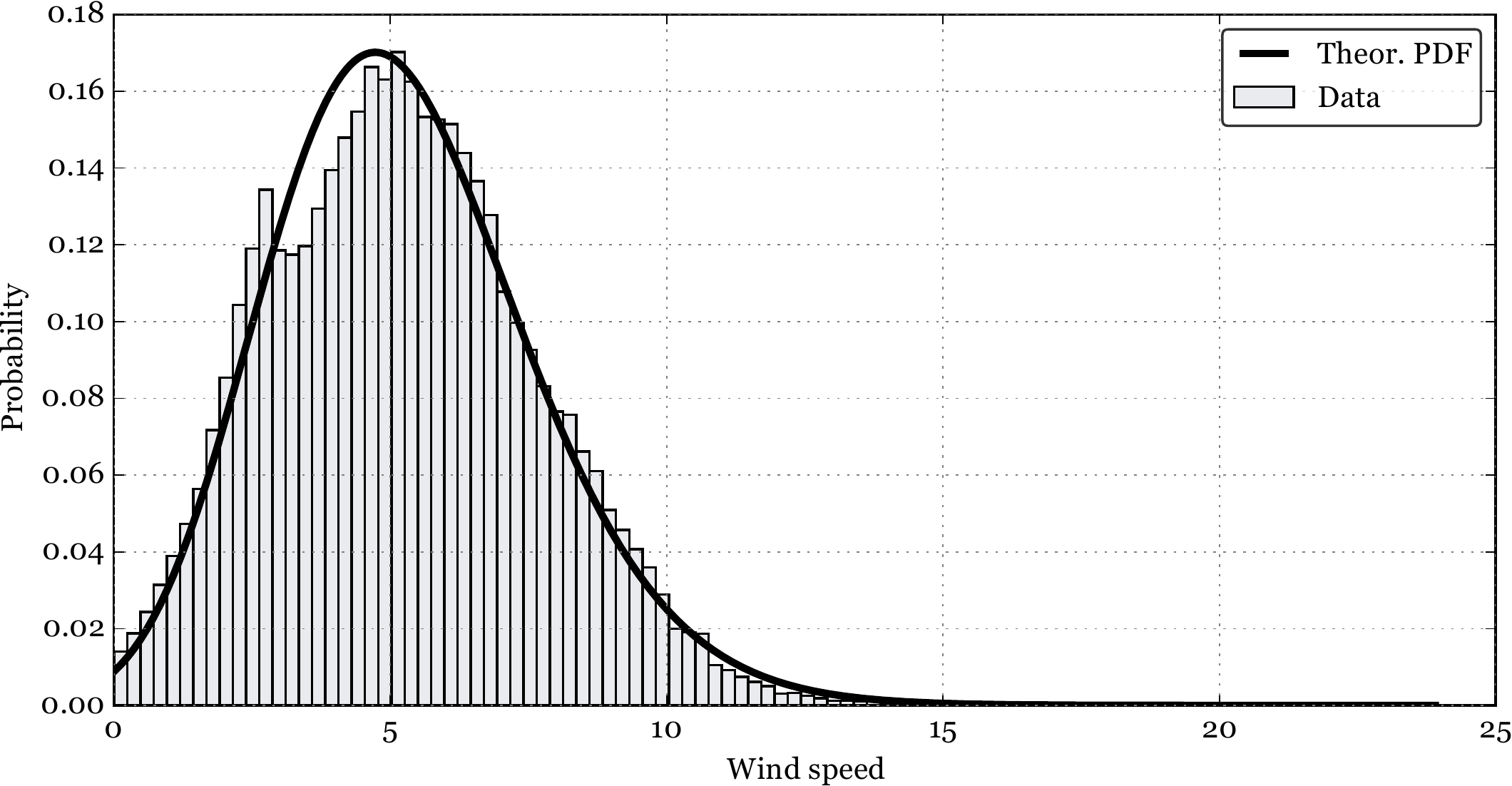}
  \caption{PDF of \textbf{Log-Normal} distribution compared with data histogram}
  \label{fig:lognormal_PDF}
\end{figure}

\begin{figure}
  \centering
  \includegraphics[width=0.8\linewidth]{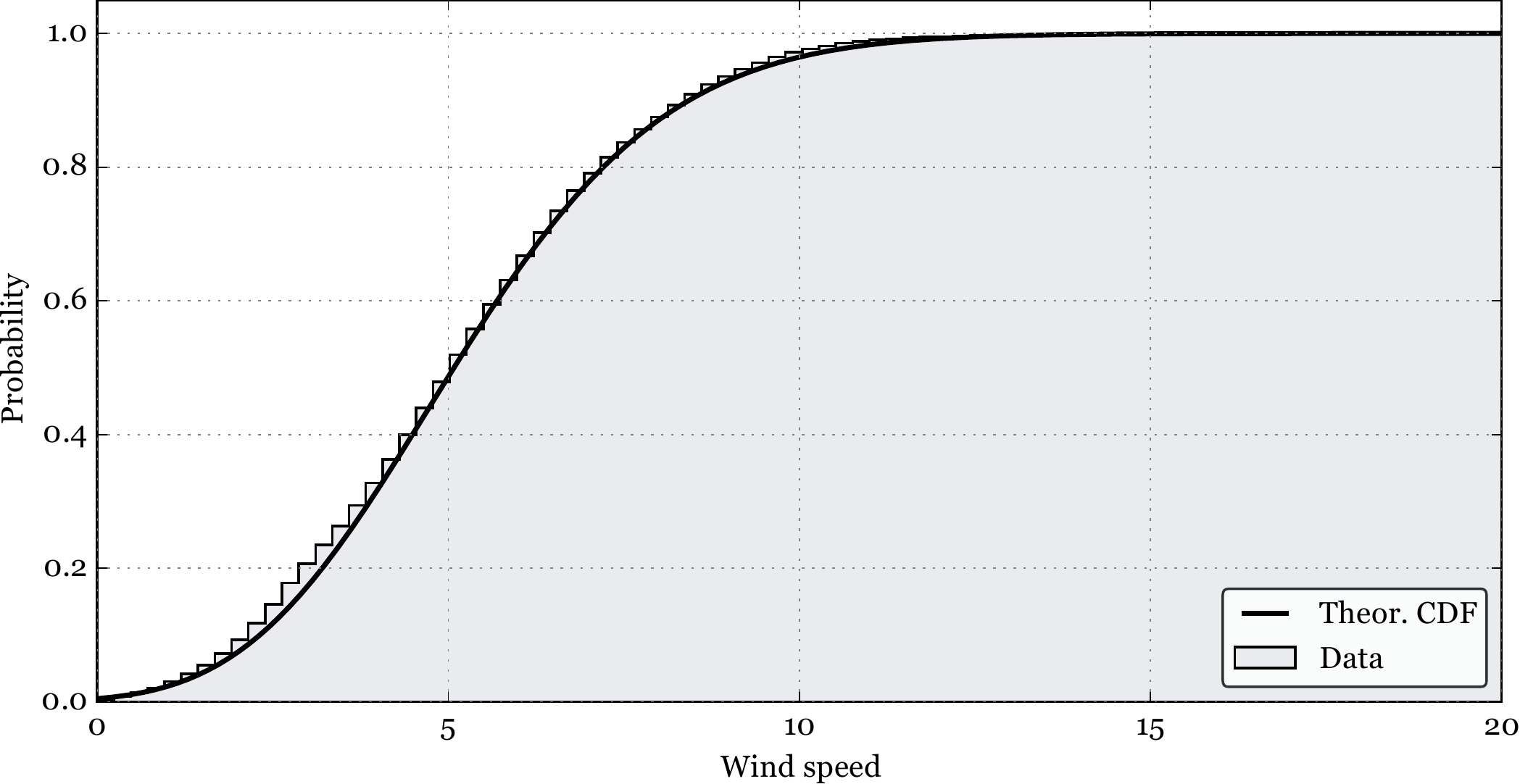}
  \caption{CDF of \textbf{Log-Normal} distribution compared with empirical distribution function}
  \label{fig:lognormal_CDF}
\end{figure}

\begin{figure}
  \centering
  \includegraphics[width=0.8\linewidth]{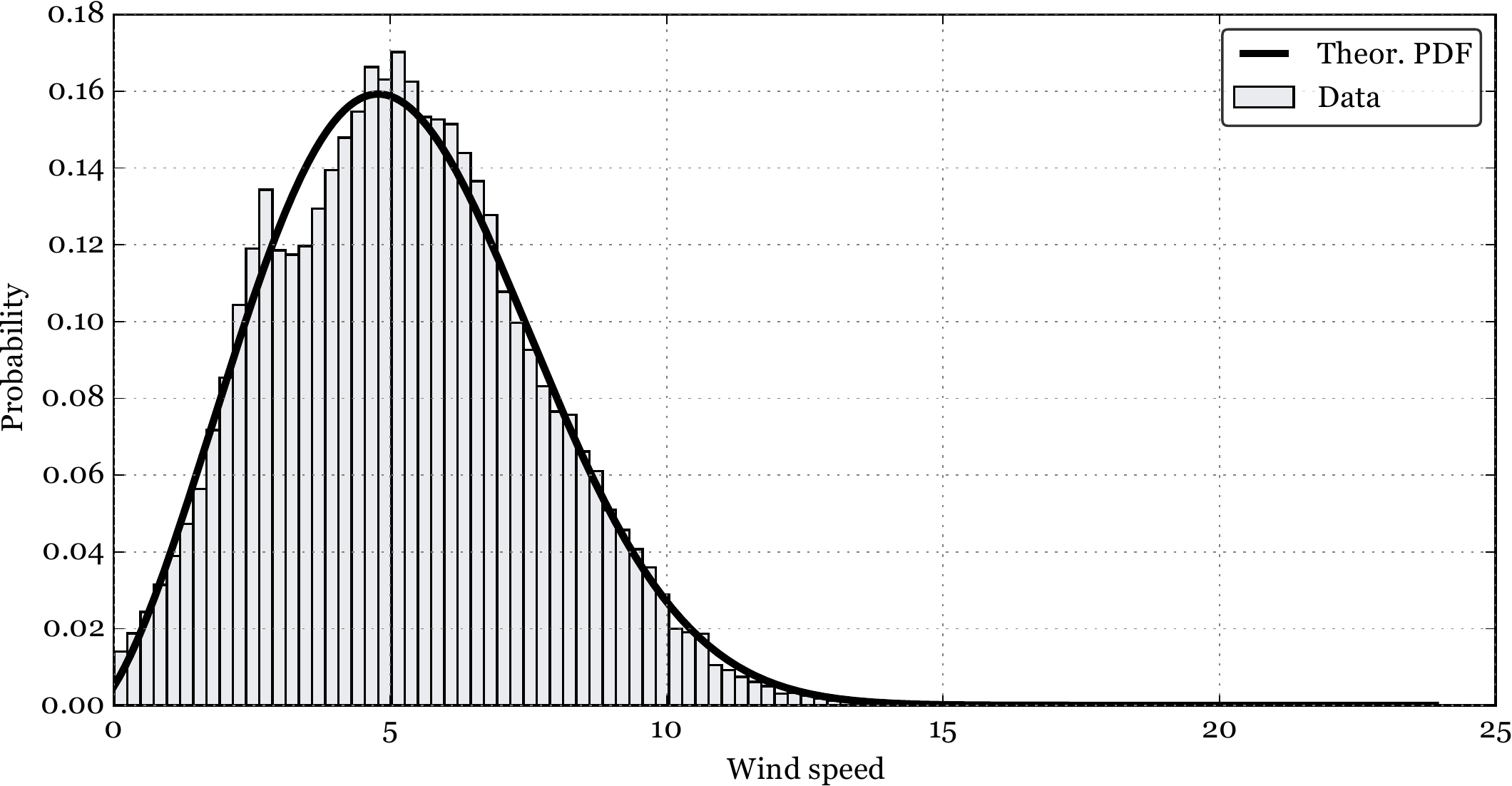}
  \caption{PDF of \textbf{Weibull} distribution compared with data histogram}
  \label{fig:Weibull_PDF}
\end{figure}

\begin{figure}
  \centering
  \includegraphics[width=0.8\linewidth]{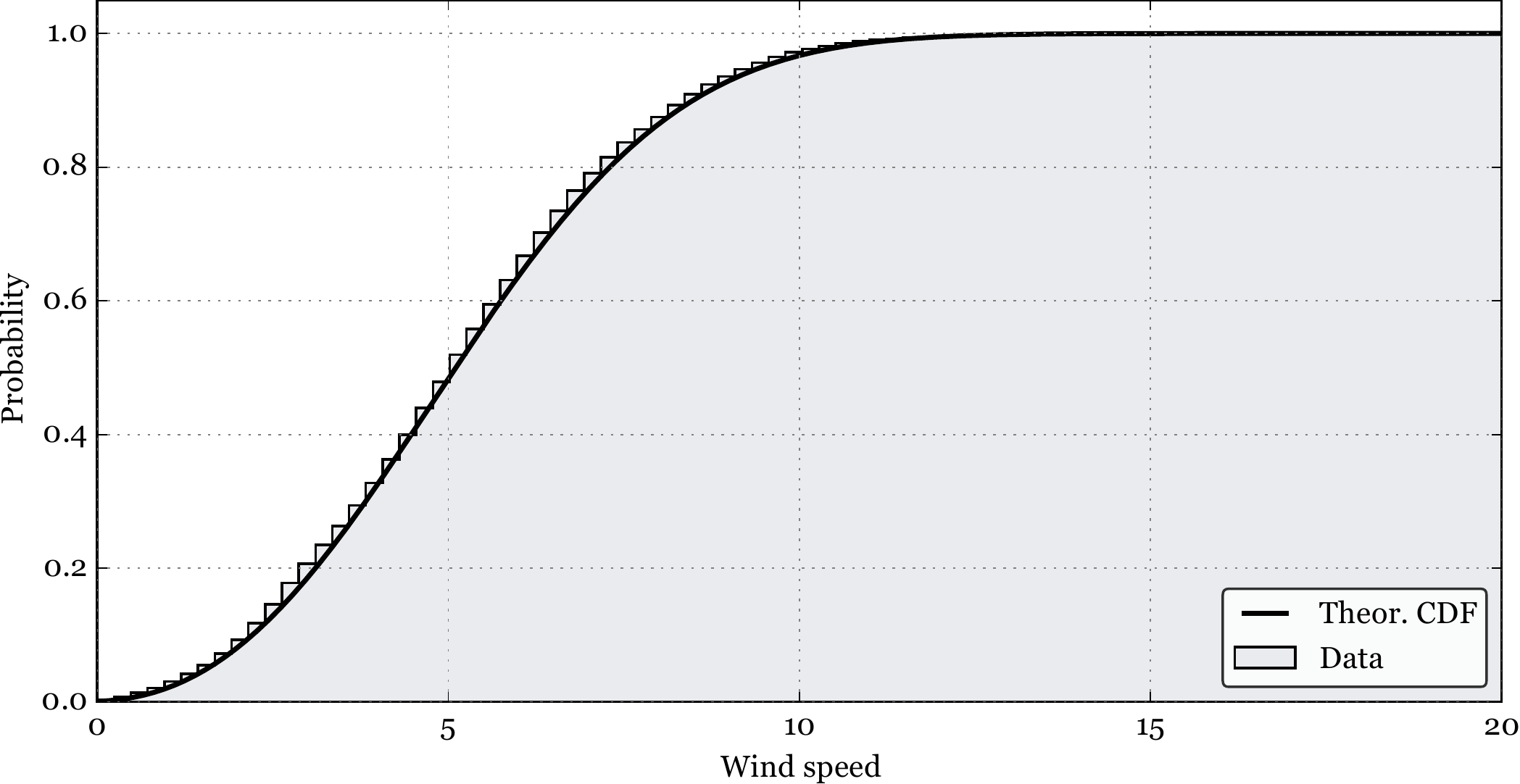}
  \caption{CDF of \textbf{Weibull} distribution compared with empirical distribution function}
  \label{fig:Weibull_CDF}
\end{figure}

\begin{figure}
  \centering
  \includegraphics[width=0.8\linewidth]{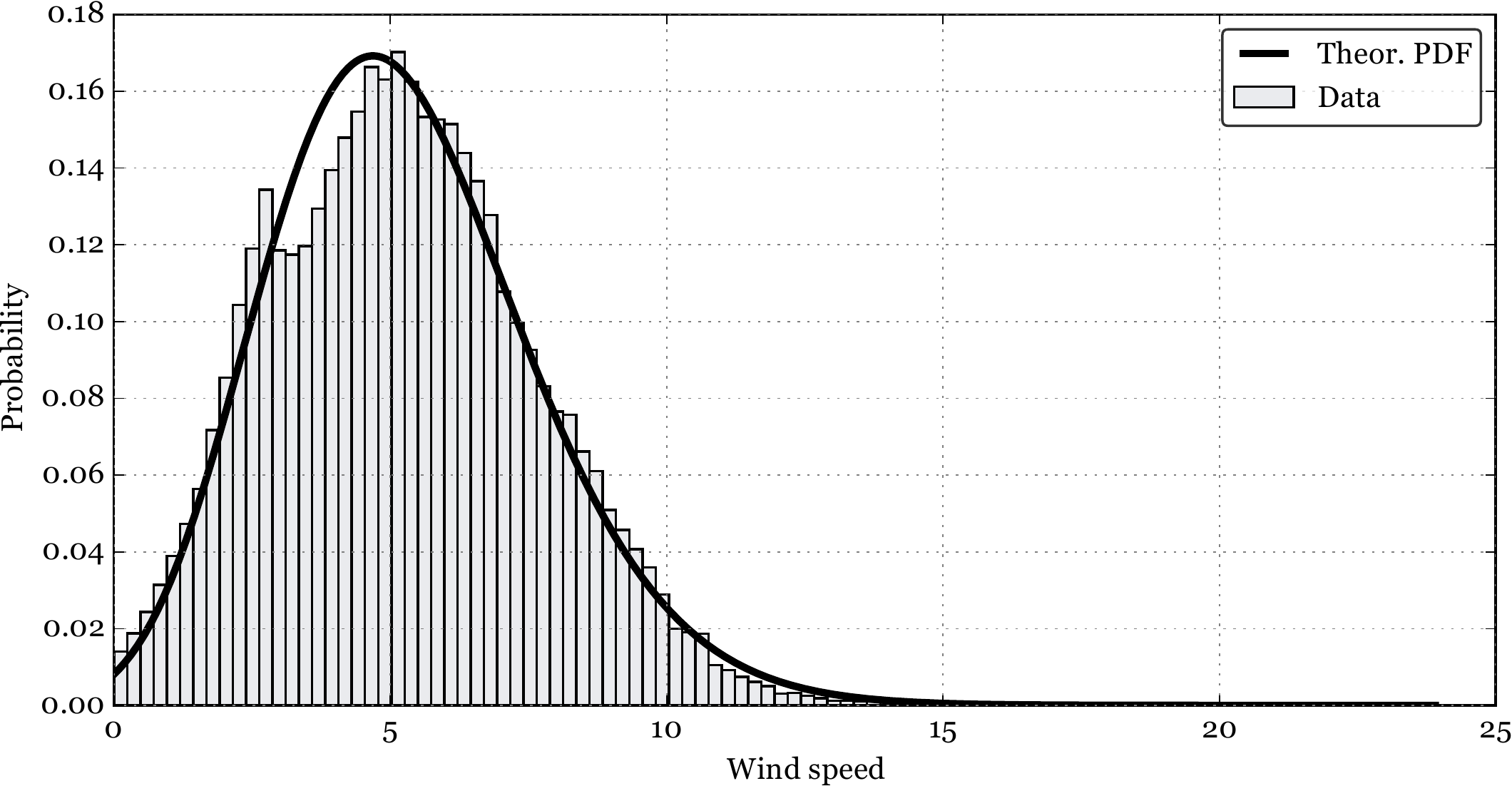}
  \caption{PDF of \textbf{Gamma} distribution compared with data histogram}
  \label{fig:Gamma_PDF}
\end{figure}

\begin{figure}
  \centering
  \includegraphics[width=0.8\linewidth]{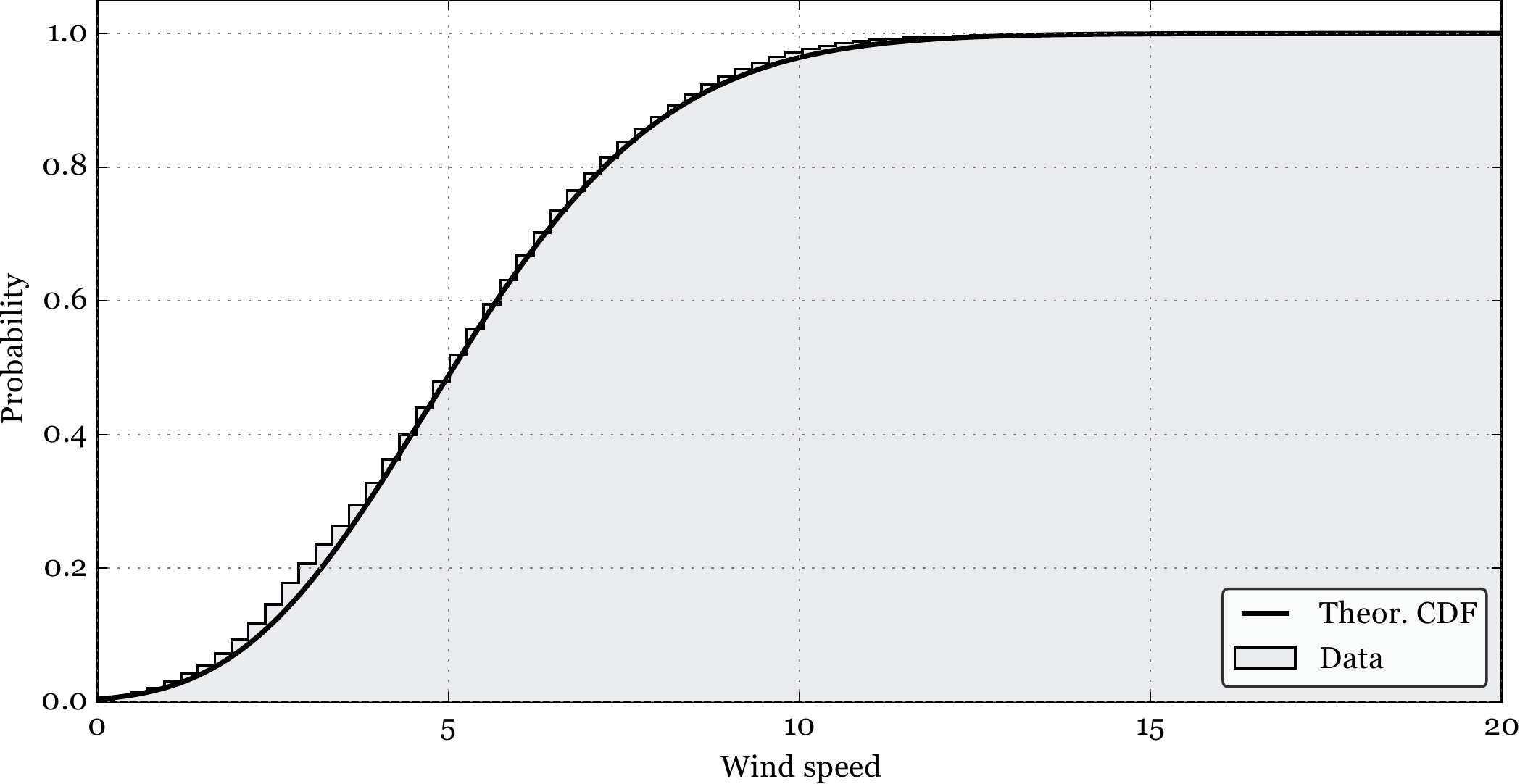}
  \caption{CDF of \textbf{Gamma} distribution compared with empirical distribution function}
  \label{fig:Gamma_CDF}
\end{figure}

\begin{figure}
  \centering
  \includegraphics[width=0.8\linewidth]{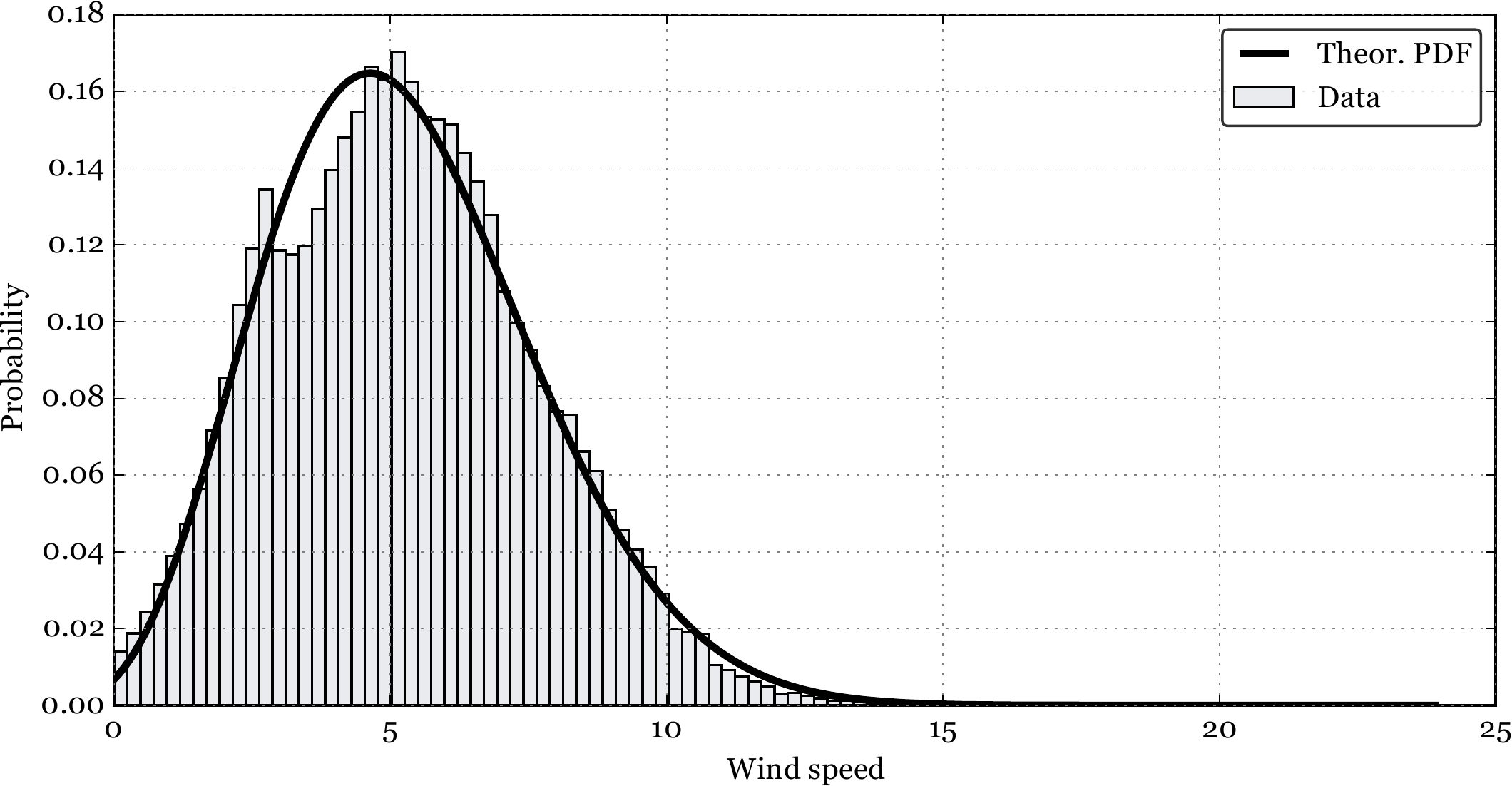}
  \caption{PDF of \textbf{Beta} distribution compared with data histograms}
  \label{fig:Beta_PDF}
\end{figure}

\begin{figure}
  \centering
  \includegraphics[width=0.8\linewidth]{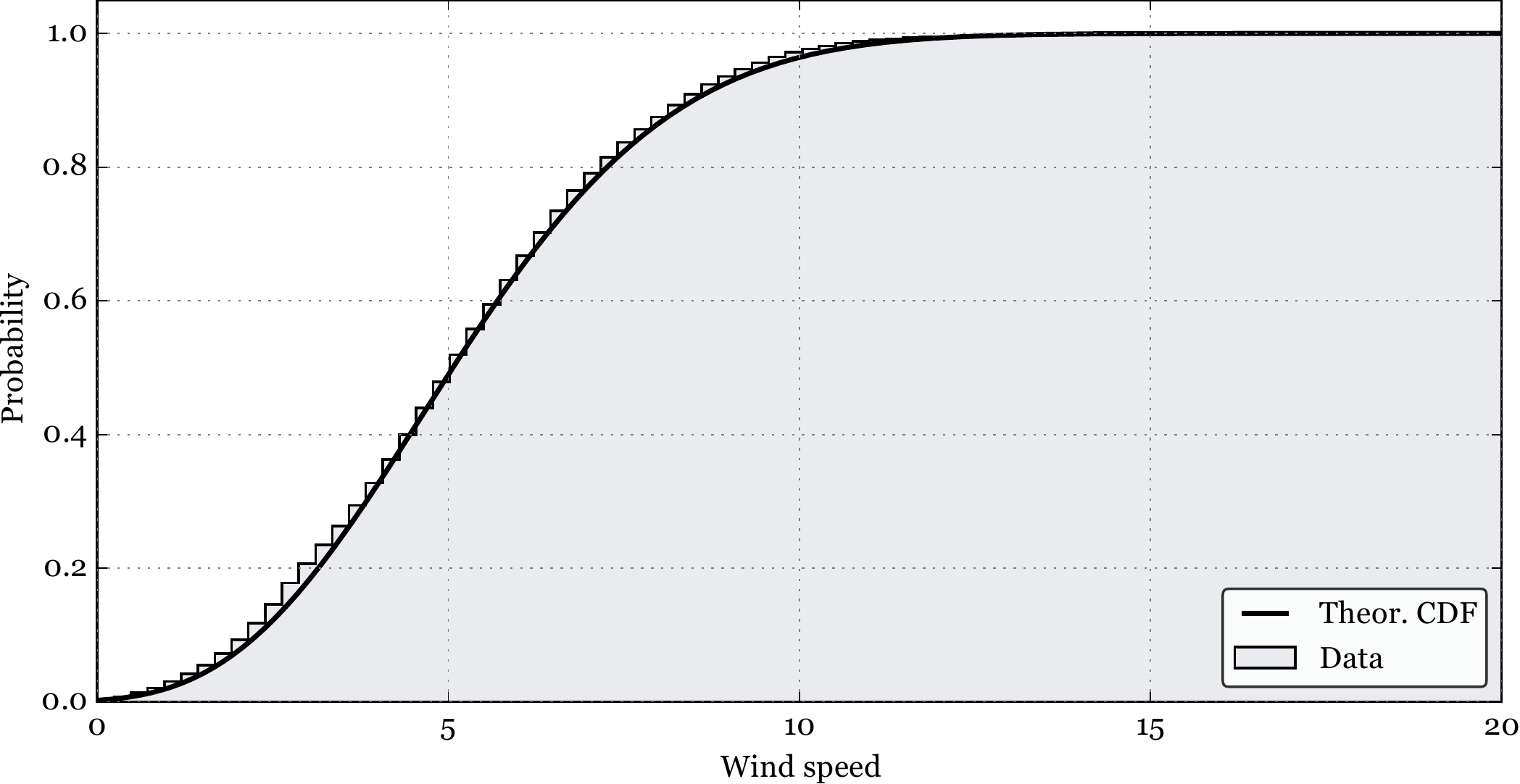}
  \caption{CDF of \textbf{Beta} distribution compared with empirical distribution function}
  \label{fig:Beta_CDF}
\end{figure}

\begin{figure}
  \centering
  \includegraphics[width=0.8\linewidth]{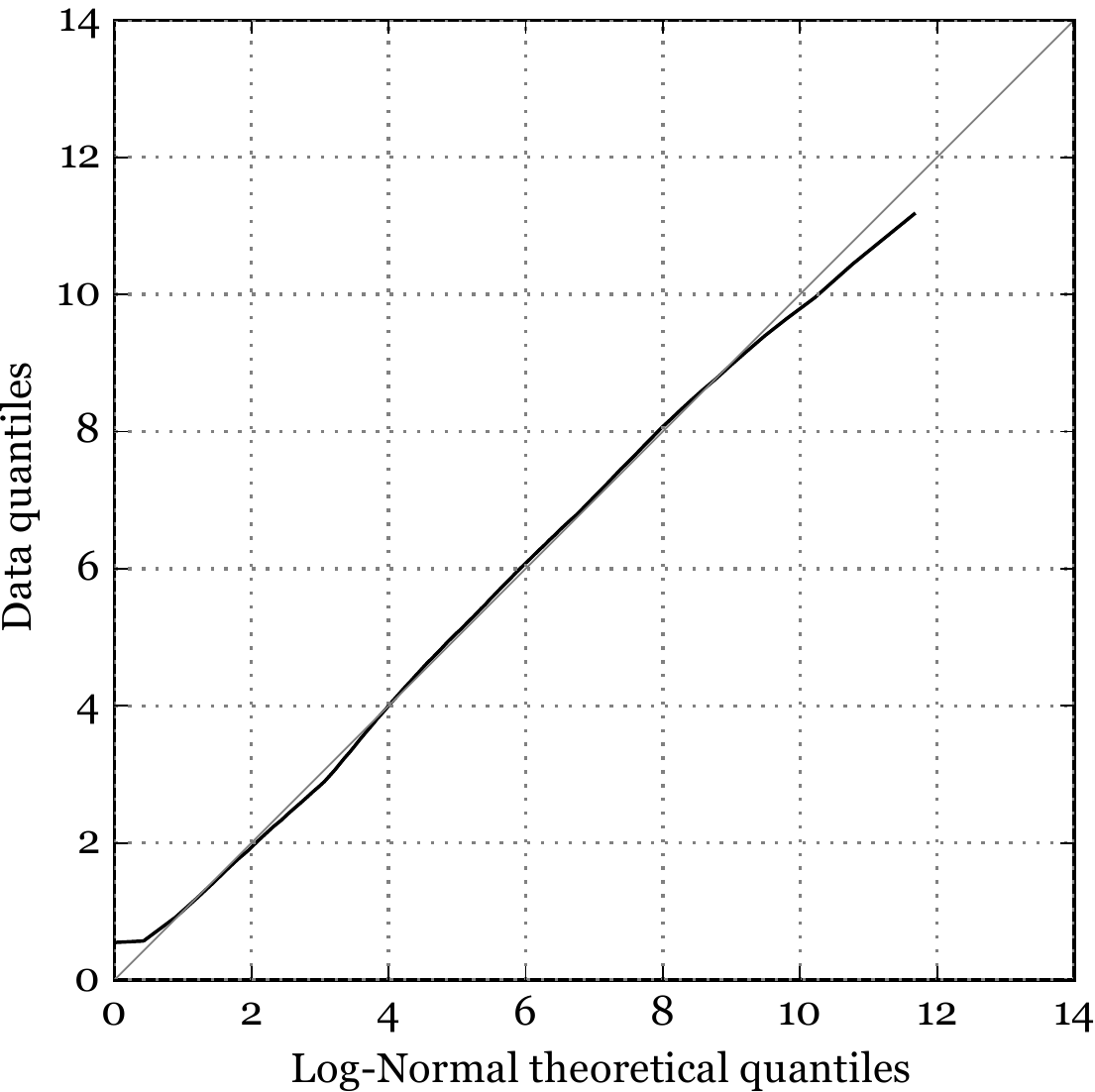}
  \caption{Q-Q plot for LogNormal distribution}
  \label{fig:LogNorm_qqplot}
\end{figure}

\begin{figure}
  \centering
  \includegraphics[width=0.8\linewidth]{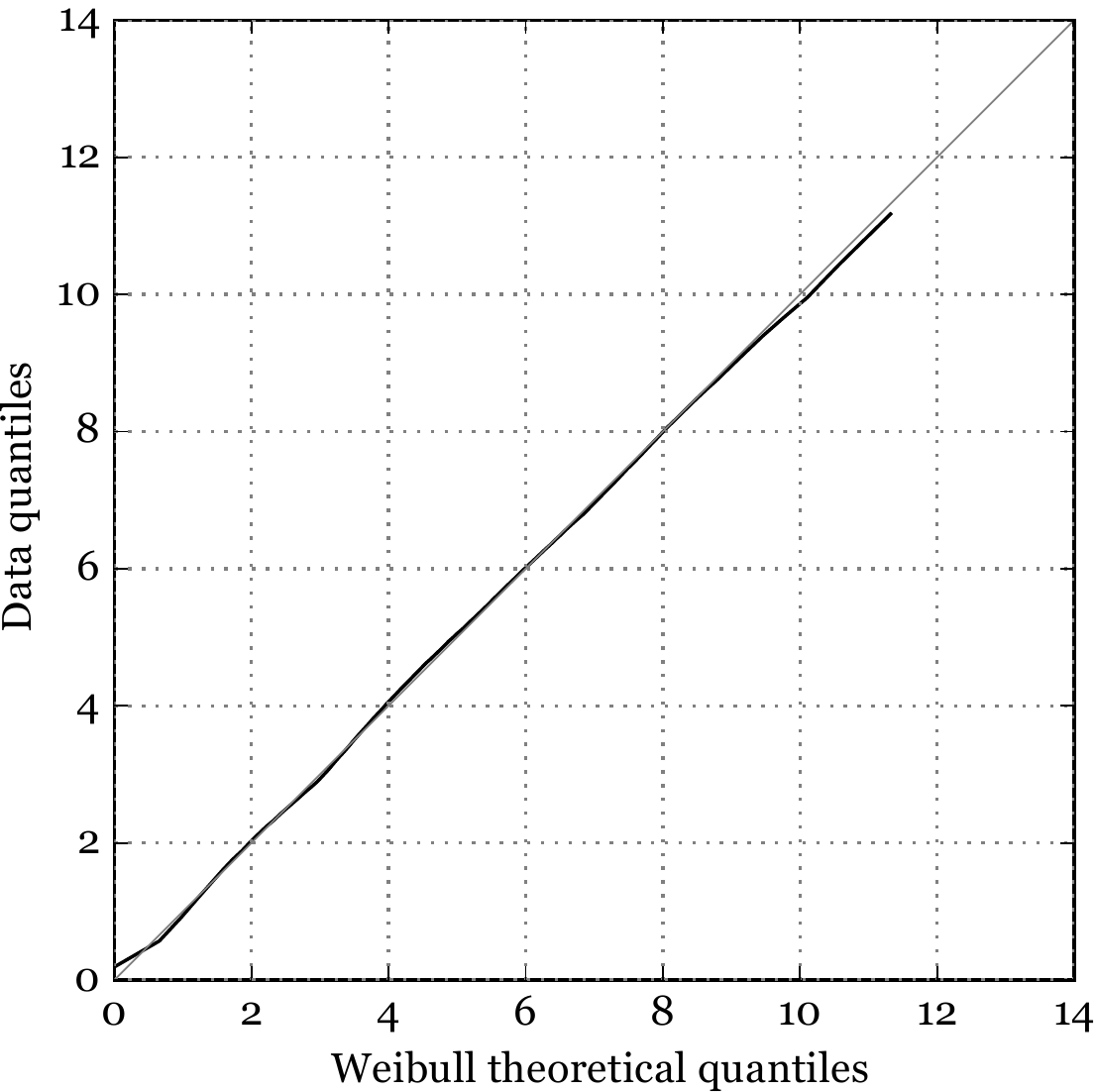}
  \caption{Q-Q plot for Weibull distribution}
  \label{fig:Weibull_qqplot}
\end{figure}

\begin{figure}
  \centering
  \includegraphics[width=0.8\linewidth]{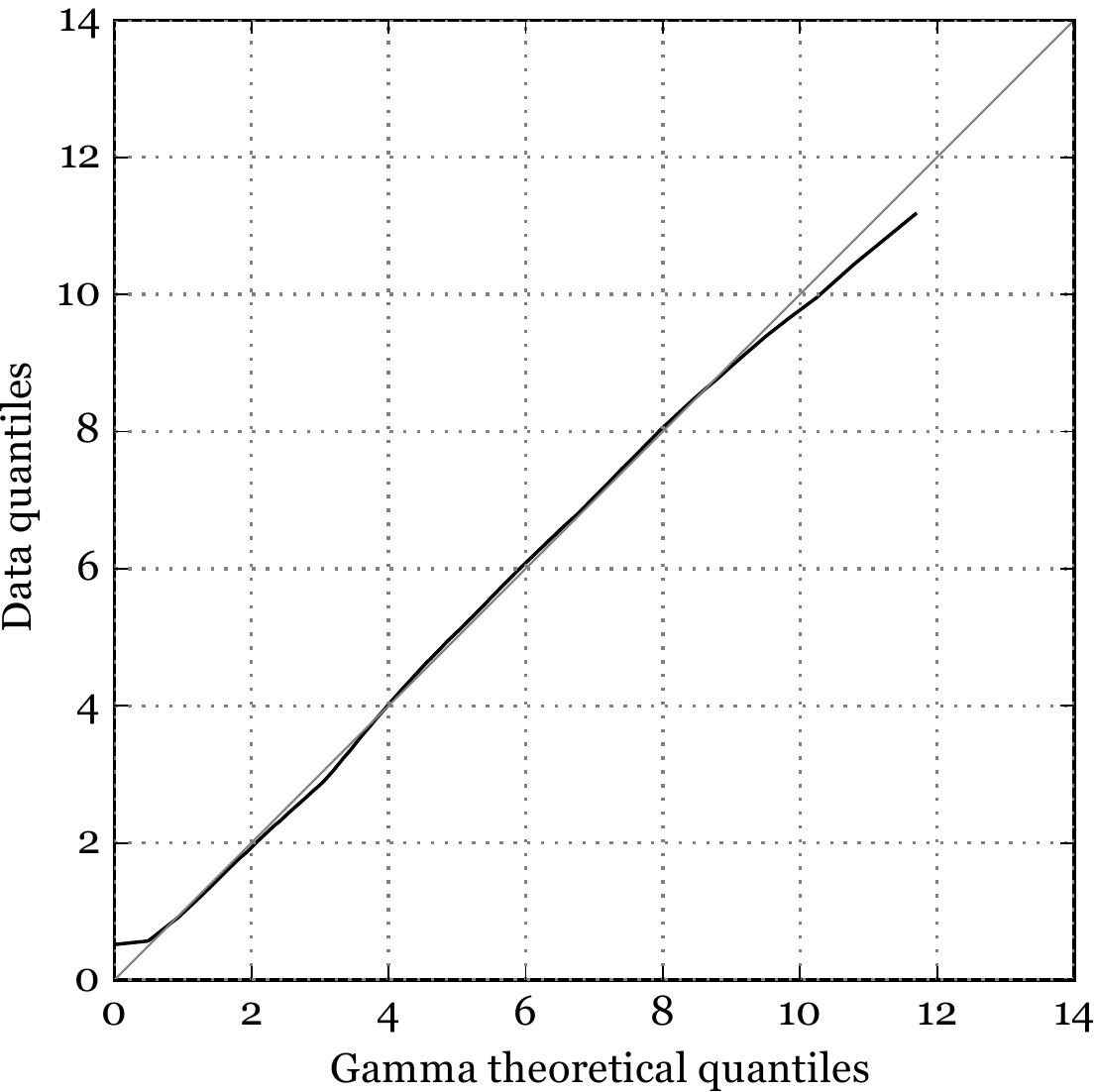}
  \caption{Q-Q plot for Gamma distribution}
  \label{fig:Gamma_qqplot}
\end{figure}

\begin{figure}
  \centering
  \includegraphics[width=0.8\linewidth]{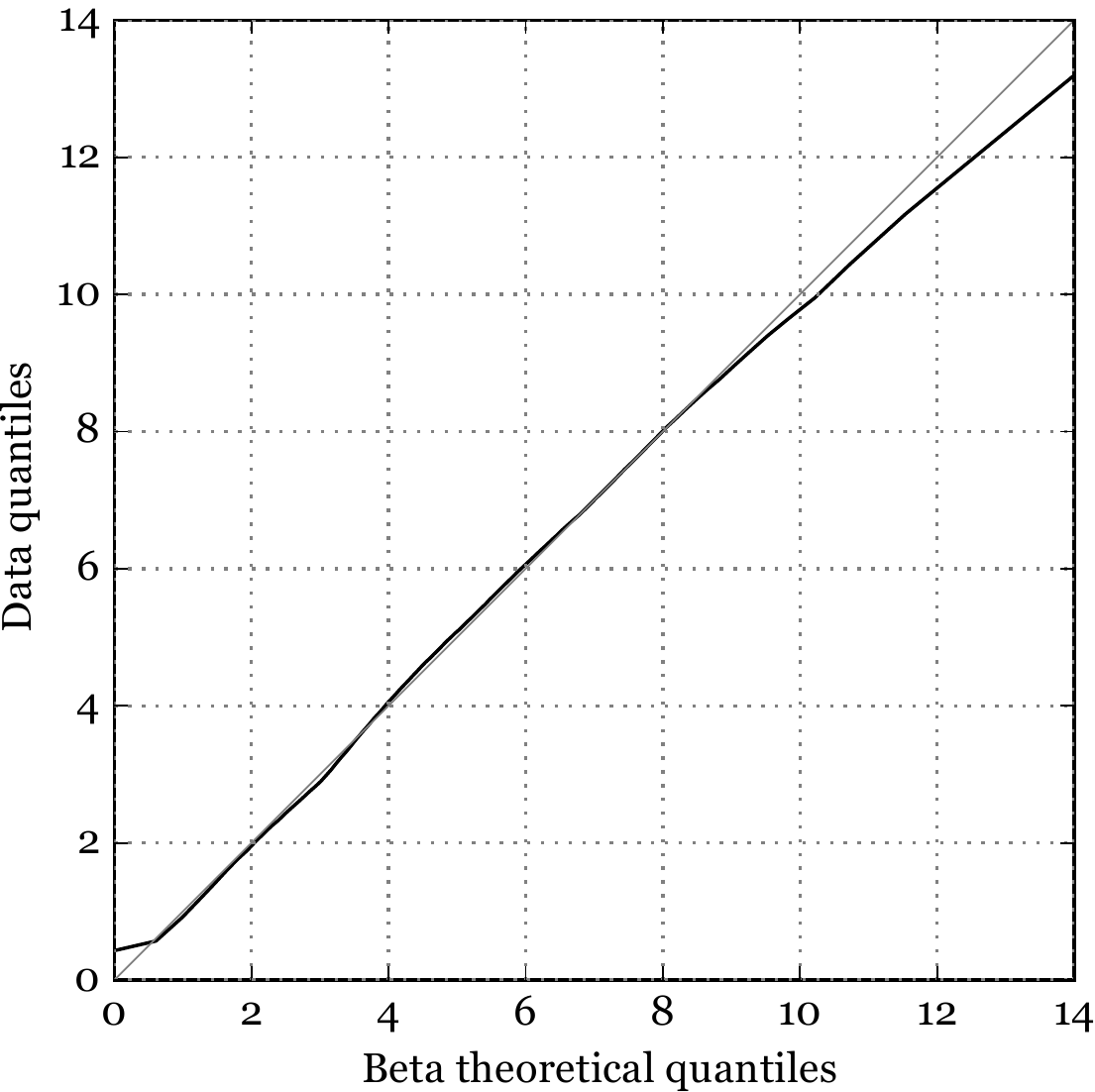}
  \caption{Q-Q plot for Beta distribution}
  \label{fig:Beta_qqplot}
\end{figure}

The results are presented graphically on figures~\ref{fig:lognormal_PDF},~\ref{fig:lognormal_CDF},~\ref{fig:Weibull_PDF},~
\ref{fig:Weibull_CDF},~\ref{fig:Gamma_PDF},~\ref{fig:Gamma_CDF},~\ref{fig:Beta_PDF},~\ref{fig:Beta_CDF} and 
\ref{fig:Weibull_qqplot},~\ref{fig:LogNorm_qqplot},~\ref{fig:Gamma_qqplot},~\ref{fig:Beta_qqplot}. The figures 
were plotted for theoretical distributions, the parameters of which have been determined 
on the basis of the entire dataset. From the analysis of the quantile-quantile plots (Q-Q plots) we can conclude 
that the Weibull distribution is best suited for approximation of available data (although only slightly), 
outmatching them only in the approximation of extreme values of a random variable.

We also performed computations with the considered distributions parameterized by only two parameters 
(let $l=0$, and for Beta distribution an addition let $s=1$). After plotting the results of calculations 
we found out that the two-parameter Weibull distribution has superiority over other two-parameters 
distributions (Log-Normal, Gamma and Beta), which is not true for three-parameter case.

\section{Conclusions}

The results of statistical data processing correspond to the results presented in the literature, where Weibull 
distribution is the most often used distribution for the wind speed approximation~\cite{L_Lun2000145,L_Seguro200075,L_1983WiEng,L_4488041,L_Islam2011985,L_Garcia1998139}.

Our future work will be aimed at the construction of stochastic models that can approximate the wind speed depending on 
time~\cite{L_1511.02345}. On the other hand, we expect to verify the results of this work by using more dilated and large data array.

\label{sec:conclusion}

\begin{acknowledgments}

The work is partially supported by RFBR grants No's 15-07-08795 and 16-07-00556.
Also the publication was supported by the Ministry of Education and
Science of the Russian Federation (the Agreement No~02.a03.21.0008).
The computations were carried out on the Felix computational cluster
(RUDN University, Moscow, Russia) and on the HybriLIT
heterogeneous cluster (Multifunctional center for data storage,
processing, and analysis at the Joint Institute for Nuclear
Research, Dubna, Russia).

\end{acknowledgments}

 \bibliographystyle{elsarticle-num}

\bibliography{bib/weibull-test/cite}

\end{document}

\graphicspath{{image/weibull-test/en/}}

\title{Исследование подходов к стохастическому моделированию ветроэнергетических установок}

\author{М. Н. Геворкян}
\email{gevorkyan_mn@rudn.university}
\affiliation{Кафедра прикладной информатики и теории вероятностей,\\
  Российский университет дружбы народов,\\
  ул. Миклухо-Маклая, д. 6, Москва, Россия, 117198}

\author{А. В. Демидова}
\email{demidova_av@rudn.university}
\affiliation{Кафедра прикладной информатики и теории вероятностей,\\
  Российский университет дружбы народов,\\
  ул. Миклухо-Маклая, д. 6, Москва, Россия, 117198}

\author{Р. А. Соболевский}
\email{r.sobolewski@pb.edu.pl}
\affiliation{Кафедра энергетики, машиностроения, фотоники и технологий освещения,\\
  Белостокский технологический университет,\\
  Республика Польша, 15-351 Белосток, ул. Wiejska 45 D}

\author{И. С. Зарядов}
\email{zaryadov_is@rudn.university}
\affiliation{Кафедра прикладной информатики и теории вероятностей,\\
  Российский университет дружбы народов,\\
  ул. Миклухо-Маклая, д. 6, Москва, Россия, 117198}
\affiliation{Институт проблем информатики ФИЦ ИУ РАН
  (ИПИ ФИЦ ИУ РАН),\\
  Москва, ул. Вавилова 44 кор.2, Россия, 119333}

\author{А. В. Королькова}
\email{korolkova_av@rudn.university}
\affiliation{Кафедра прикладной информатики и теории вероятностей,\\
  Российский университет дружбы народов,\\
  ул. Миклухо-Маклая, д. 6, Москва, Россия, 117198}

\author{Д. С. Кулябов}
\email{kulyabov_ds@rudn.university}
\affiliation{Кафедра прикладной информатики и теории вероятностей,\\
  Российский университет дружбы народов,\\
  ул. Миклухо-Маклая, д. 6, Москва, Россия, 117198}
\affiliation{Лаборатория информационных технологий,\\
  Объединённый институт ядерных исследований,\\
  ул. Жолио-Кюри 6, Дубна, Московская область, Россия, 141980}

\author{Л. А. Севастьянов}
\email{sevastianov_la@rudn.university}
\affiliation{Кафедра прикладной информатики и теории вероятностей,\\
  Российский университет дружбы народов,\\
  ул. Миклухо-Маклая, д. 6, Москва, Россия, 117198}
\affiliation{Лаборатория теоретической физики,\\
  Объединённый институт ядерных исследований,\\
  ул. Жолио-Кюри 6, Дубна, Московская область, Россия, 141980}

\begin{abstract}

\textbf{Предпосылки.} В работе исследуются статистические данные, собранные с ветроэнергетических установок, 
находящихся на территории Республики Польши. исследования направленны на построение стохастической модели, 
предсказывающей изменение скорости ветра в зависимости от времени.
\textbf{Целью} данной работы является поиск наиболее оптимального распределения для аппроксимации имеющихся 
статистических данных по скорости ветра.
\textbf{Методы} Рассматриваются четыре распределения случайной величины: логнормальное, Вейбулла, гамма и бета. 
Из массива статистических данных методом функции максимального правдоподобия находятся параметры данных распределений. 
Для оценки аппроксимации используется квантиль-кванлиль (Q-Q) график. 
\textbf{Результаты} Все рассматриваемые распределения  имеющиеся данные. Распределение 
Вейбулла показывает лучшие результаты аппроксимации для крайних значений скорости ветра.
\textbf{Выводы} Результаты анализа согласуются с общепринятой практикой использования распределения 
Вейбулла для моделирования скорости ветра. В дальнейшем планируется сравнить полученные результаты с 
гораздо большим массивом данных, а также построить стохастическую модель эволюции скорости ветра в зависимости от времени.

\end{abstract}

  \keywords{аппроксимация, распределение Вейбулла, логнормальное
    распределение, гамма распределение, бета распределение, скорость
    ветра, статистические данные}

\maketitle

\section{Введение}
Данная работа посвящена проблеме стохастического моделирования изменения 
скорости ветра, используемого для генерации электроэнергии на 
ветроэнергетических установках находящихся на территории Республики 
Польши (Republic of Poland). В качестве первого этапа выбирается вид 
теоретического распределения случайной величины $w$, которая должна 
аппроксимировать скорость ветра.

В данной работе мы исследовали несколько распределений на предмет 
аппроксимации располагаемых нами данных, а именно: логнормальное, 
Вейбулла, а также бета и гамма распределения. Все рассматриваемые 
распределения имеют параметризацию коэффициентами формы, сдвига и 
масштаба (shape-location-scale parametrisation). Для обработки 
статистических данных применялся язык Python 3 в связке с библиотеками 
\texttt{numpy}, \texttt{scipy.stats}~\cite{L_scipy} и 
\texttt{matplotlib}~\cite{L_matplotlib}, а также интерактивной 
оболочкой \texttt{Jupyter}~\cite{L_jupyter}.

Мы использовали книги~\cite{L_Johnson1_en,L_Johnson2_en,L_Nelson} в качестве справочных материалов по свойствам используемых нами распределений. Статьи~\cite{L_Frechet,L_Weibull} --- первоисточники, в которых распределение Вейбулла представлено впервые. Статьи~\cite{L_Lun2000145,L_Seguro200075,L_1983WiEng,L_4488041,L_Islam2011985,L_Garcia1998139} описывают использование распределения Вейбулла при моделировании ветроэнергетических установок и скорости ветра.
\label{sec:intro}

\section{Описание структуры статистических данных}

Статистические данные представляют собой таблицу величин в формате \texttt{csv} составленную из следующих колонок:
\begin{enumerate}
  \item $T$ — время фиксации скорости и направления ветра датчиками, установленными на ветроэнергетической турбине (в формате чч:мм);
  \item $X_1$ — мощность турбины [кВт] (отрицательные значения означают, что электроэнергия не производится, а потребляется);
  \item $X_2$ — скорость ветра на  [м/с] (измеряется анемометром на вершине гондолы воздушной турбины),
  \item $X_3$ — направление ветра [градусы] (измеряется анемометром на вершине гондолы воздушной турбины; измеряется по часовой стрелке, 
  значения от $0^{\circ}$ до $N^{\circ}$);
  \item $X_4$ — скорость ветра на уровне 10 метров от поверхности земли [м/с];
  \item $X_5$ — направление ветра на уровне 10 метров от поверхности земли [градусы];
  \item $X_6$ — скорость ветра на уровне 50 метров [м/с] от поверхности земли;
  \item $X_7$ — направление ветра на уровне 50 метров от поверхности земли [градусы].
\end{enumerate}
Показатели скорости и направления ветра снимались с датчиков каждые 10 минут на протяжении около 9 месяцев. 
В общей сложности таблица содержит 39606 записей.  

\begin{figure}
  \centering
  \includegraphics[width=0.9\linewidth]{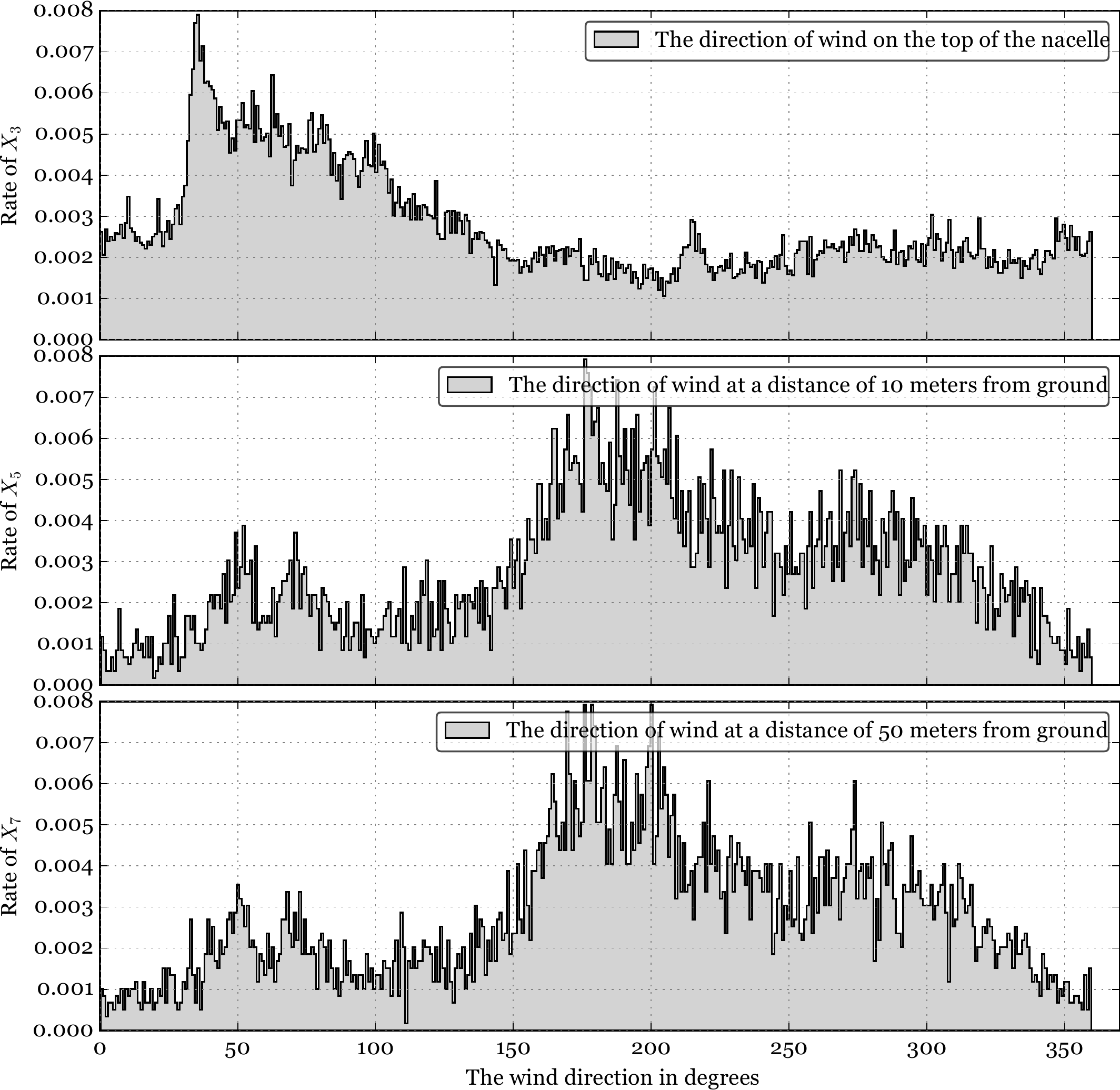}
  \caption{Histogram of wind direction at three levels of height}
  \label{fig:winddir}
\end{figure}

Чтобы осуществить первоначальный выбор распределений, которые могут
подойти для аппроксимации скорости ветра мы нарисовали гистограммы
распределения скоростей ветра. Визуальная оценка этих гистограмм
позволила предположить, что адекватным выбором будут распределения с
«тяжелыми хвостами». Для аппроксимации направлений ветра такие
распределения не подойдут, в чем можно убедиться посмотрев на
гистограмму на рис.~\ref{fig:winddir}.

Для считывания данных мы использовали функцию \texttt{genfromtxt}
библиотеки \texttt{numpy}~\cite{L_scipy}
\lstset{language=Python,breaklines=true,breakatwhitespace=true}
\begin{lstlisting}
ws1, ws2, ws3 = np.genfromtxt('data.csv', delimiter=';', skip_header=True, usecols=(2, 4, 6), unpack=True)
\end{lstlisting}
где \texttt{'data.csv'} --- файл с данными, \texttt{delimiter=';'} ---
разделитель колонок, \texttt{skip\_header = True} --- указывает на
необходимость игнорирования первой строки, так как в ней перечислены
названия колонок, \texttt{usecols=(2, 4, 6)} --- использовать только
2,4,6 колонки (нумерация начинается с нуля) и \texttt{unpack=True} ---
содержимое каждой из колонок следует записать в отдельные массивы
\texttt{ws1}, \texttt{ws2} и \texttt{ws3} для дальнейшего анализа
данных по отдельности.

  \section{Используемые распределения}

Каждое из рассматриваемых в работе распределений параметризовано тремя
параметрами: $\alpha$ --- коэффициент формы (shape), $l$ ---
коэффициент сдвига (location) и $s$ --- коэффициент масштаба
(scale). В случае бета-распределения добавляется еще второй
коэффициент масштаба, обозначаемый буквой $\beta$. Для всех
распределений параметры полагаются положительными действительными
числами: $\alpha, \beta, s, l \in \mathbb{R}$, $\alpha, \beta, s > 0$,
$l\geqslant 0$. Напомним основные определения.

Случайная величина $X$ распределена по \emph{логнормальному закону},
если функция плотности вероятности задается следующей формулой:
\begin{multline}
  f_{LN}(x; \alpha, l, s)
  = {} \\ {} =
  \begin{dcases}
    \frac{1}{(x-l)\alpha \sqrt{2\pi}}\exp\left(-\frac{1}{2} \left(
        \frac{\ln(x-l) - \ln{s}}{\alpha} \right)^2 \right), & x
    \geqslant l. \\
    0, & x < l.
  \end{dcases}
\end{multline}

Случайная величина $X$ распределена по
\emph{Вейбуллу}~\cite{L_Frechet,L_Weibull}, если функция плотности
вероятности задается следующей формулой:
\begin{multline}
  f_{W}(x; \alpha, l, s)
  = {} \\ {} =
  \begin{dcases}
    \frac{\alpha}{s}\left(\frac{x-l}{s}\right)^{\alpha - 1} \exp\left[-\left(\frac{x-l}{s}\right)^{\alpha}\right], & x \geqslant l.\\
    0, & x < l.
  \end{dcases}
\end{multline}

Случайная величина $X$ подчинена \emph{гамма-распределению}, если функция плотности вероятности задается следующей формулой:
\begin{equation}
  f_{\Gamma}(x; \alpha, l, s) = 
  \begin{dcases}
    \frac{(x-l)^{\alpha-1}\exp\left(-\frac{(x-l)}{s}\right)}{s^{\alpha}\Gamma(\alpha)}, & x \geqslant l.\\
    0, & x < l.
  \end{dcases}
\end{equation}
где $\Gamma(\alpha)$ --- гамма-функция.

Случайная величина $X$ подчинена \emph{бета-распределению}, если функция плотности вероятности задается следующей формулой:
\begin{multline}
  f_{\mathcal{B}}(x; \alpha, \beta, l, s)
  = {} \\ {} =
  \begin{dcases}
    \frac{\Gamma(\alpha + \beta)}{s\Gamma(\alpha)\Gamma(\beta)}\left(\dfrac{x-l}{s}\right)^{\alpha - 1}\left(1 - 
        \frac{x - l}{s}\right)^{\beta - 1}, & x \geqslant l.\\
    0, & x < l.
  \end{dcases}
\end{multline}

Если в формулах плотности вероятности логнормального, Вейбулла и
гамма-распределений положить $l=0$, а у бета распределения в добавок
$s=1$, то мы получим вид данных распределений, наиболее часто
используемый в литературе~\cite{L_Johnson1_en,L_Nelson}.

\section{Определение параметров распределений}

В библиотеке \texttt{scipy.stats}~\cite{L_scipy} определены объекты
\texttt{lognorm}, \texttt{weibull\_min}, \texttt{gamma} и
\texttt{beta} которые реализуют рассматриваемые нами распределения. У
каждого из данных объектов предусмотрены методы, которые реализуют
функцию плотности вероятности \texttt{pdf(x, a, [b,] loc, scale)} и
функцию распределения вероятности \texttt{cdf(x, a, [b,] loc, scale)},
где \texttt{x} --- аргумент функции, \texttt{a}, \texttt{b} ---
параметры формы $\alpha$, (и $\beta$ в случае бета-распределения),
\texttt{loc} и \texttt{scale} --- параметры смещения и масштаба.

Для нахождения параметров рассматриваемых распределений в
\texttt{scipy.stats} предусмотрена функция \texttt{fit(data)}, которая
вычисляет методом максимального правдоподобия параметры
соответствующего распределения. Мы использовали данную функцию для
вычисления параметров рассматриваемых нами распределений. После чего с
помощью функций \texttt{pdf} и \texttt{cdf} были вычислены значения
функции плотности вероятности и функции распределения.

Приведём пример кода для случая логнормального закона.

\begin{lstlisting}
s, loc, scale = scipy.stats.lognorm.fit(ws1)
xs = np.linspace(np.min(ws1), np.max(ws1), 1000)
logN_PDF = scipy.stats.lognorm.pdf(xs, s, loc, scale)
logN_CDF = scipy.stats.lognorm.cdf(xs, s, loc, scale)
\end{lstlisting}
    % \hat{\mu} = \dfrac{1}{n} \sum\limits^{N}_{k=1}\ln X_{k},\;\; \hat{\sigma}^{2} = \dfrac{1}{n}\sum^{N}_{k=1}(\ln X_{k} - \hat{\mu})^2

  % \hat{k}^{-1} = \dfrac{\sum\limits_{i=1}^n x_{i}^{\hat{k}} \ln{x_i}}{\sum\limits_{i=1}^n x_{i}^{\hat{k}}} - \dfrac{1}{n} \sum\limits_{i=1}^n \ln{x_i}

\section{Результаты}

\begin{figure}
  \centering
  \includegraphics[width=0.8\linewidth]{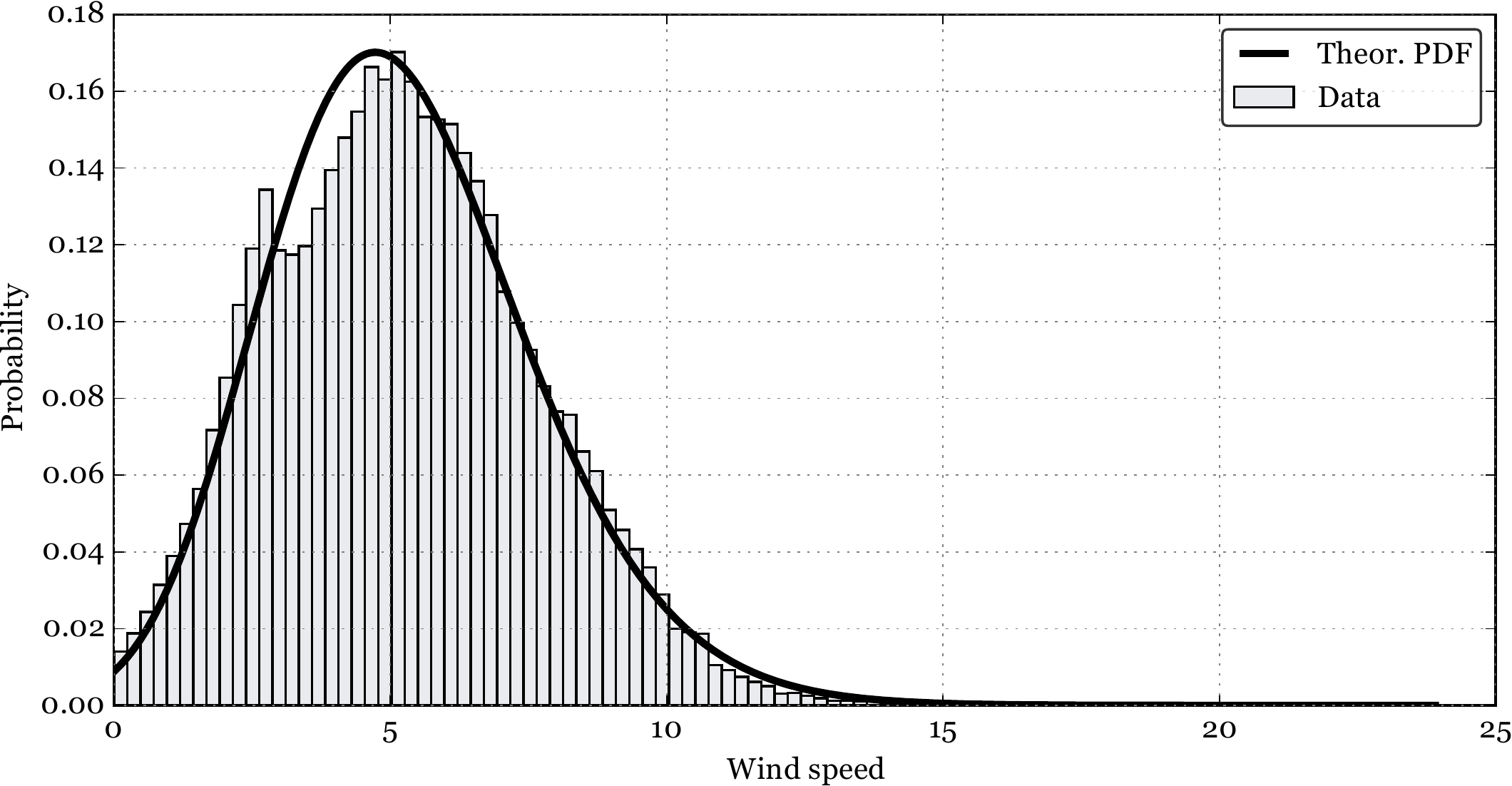}
  \caption{PDF of \textbf{Log-Normal} distribution compared with data histogram}
  \label{fig:lognormal_PDF}
\end{figure}

\begin{figure}
  \centering
  \includegraphics[width=0.8\linewidth]{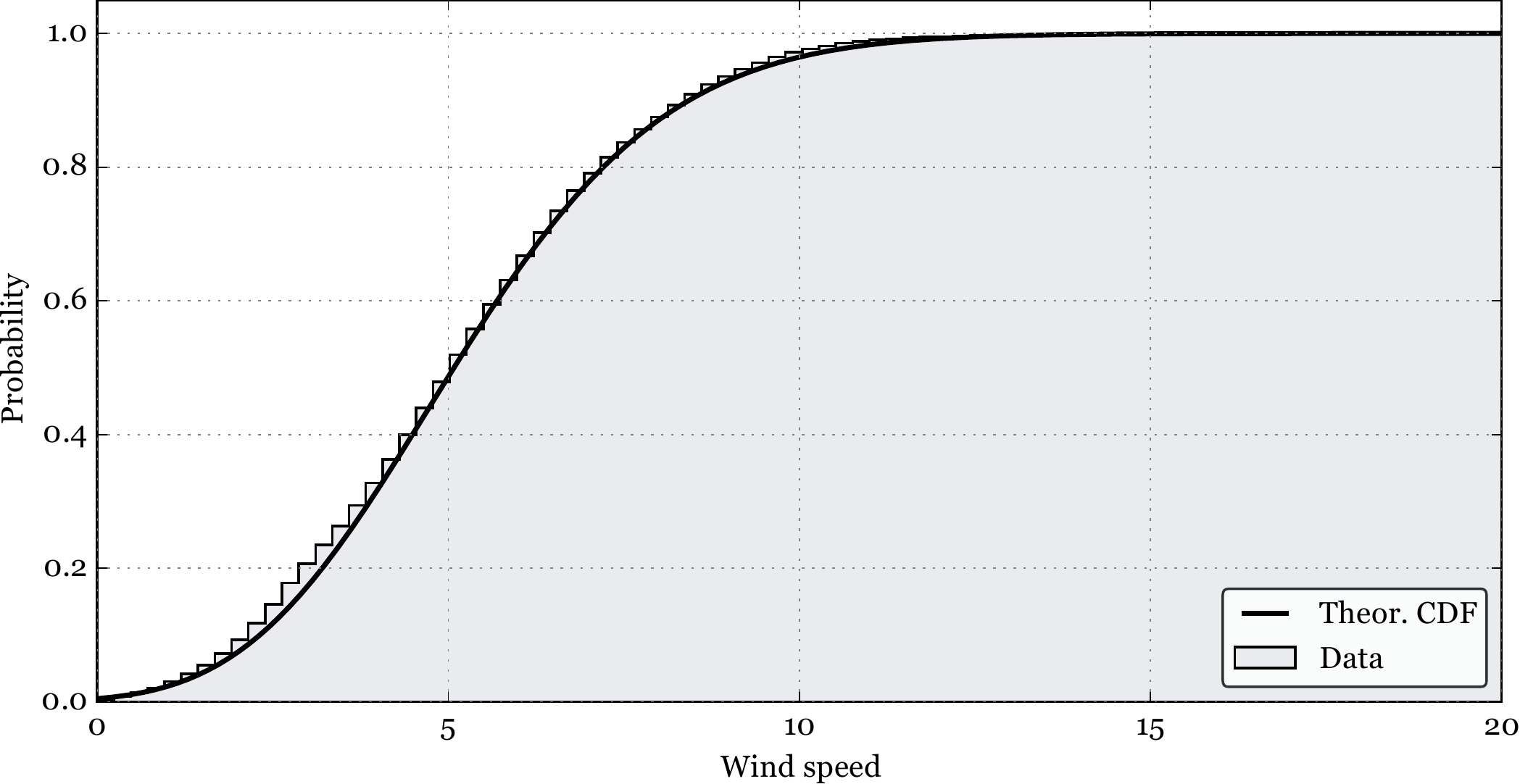}
  \caption{CDF of \textbf{Log-Normal} distribution compared with empirical distribution function}
  \label{fig:lognormal_CDF}
\end{figure}

\begin{figure}
  \centering
  \includegraphics[width=0.8\linewidth]{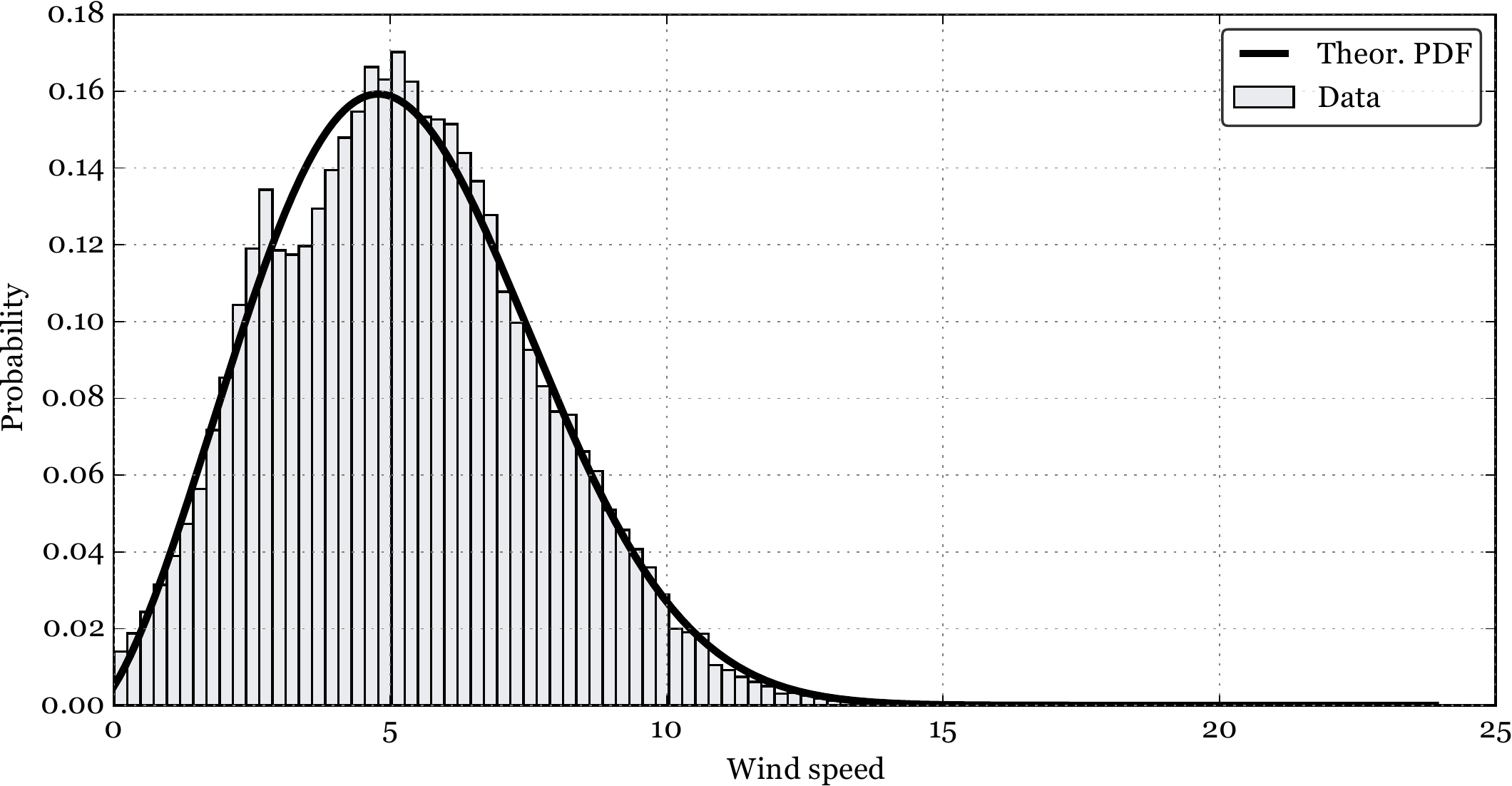}
  \caption{PDF of \textbf{Weibull} distribution compared with data histogram}
  \label{fig:Weibull_PDF}
\end{figure}

\begin{figure}
  \centering
  \includegraphics[width=0.8\linewidth]{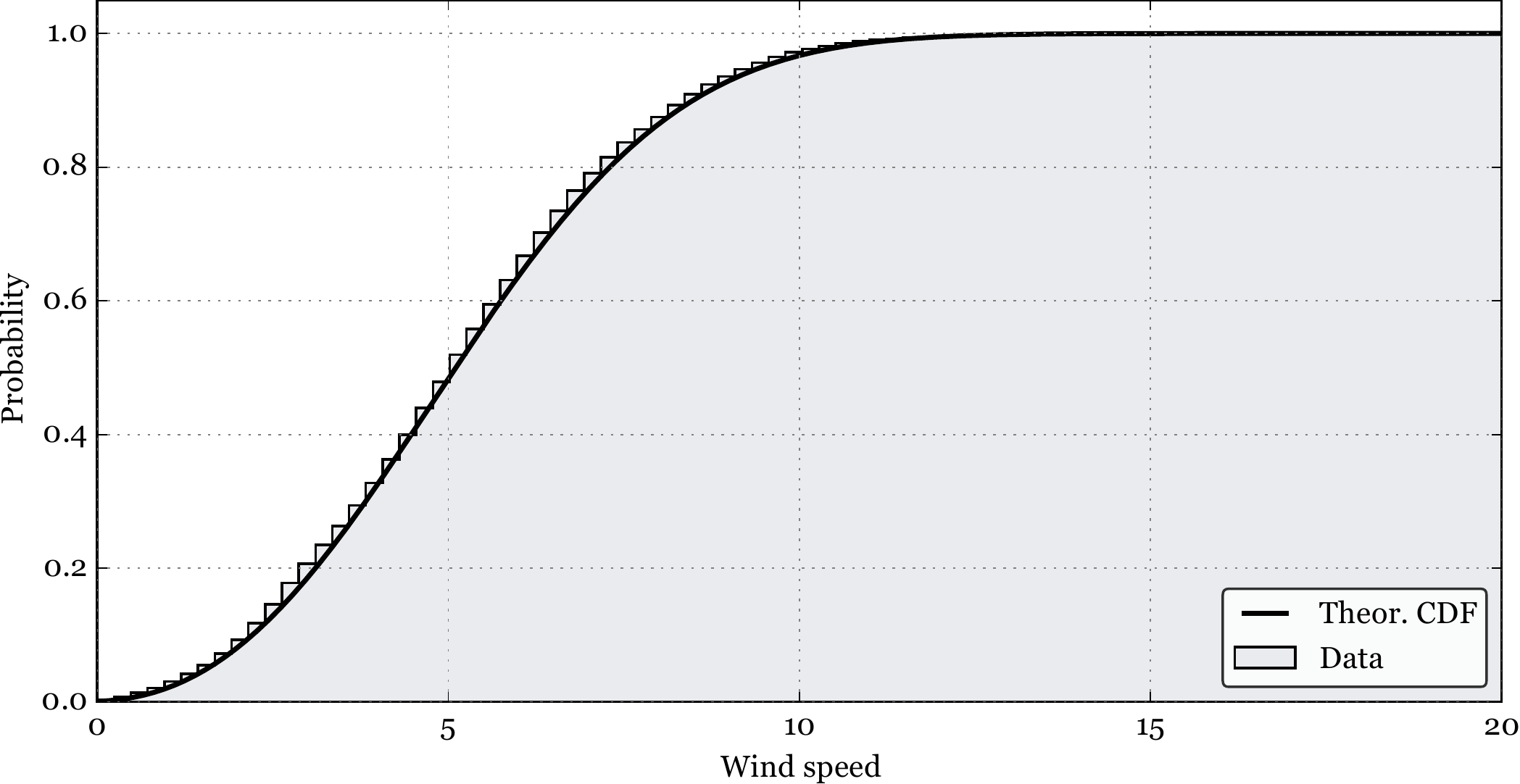}
  \caption{CDF of \textbf{Weibull} distribution compared with empirical distribution function}
  \label{fig:Weibull_CDF}
\end{figure}

\begin{figure}
  \centering
  \includegraphics[width=0.8\linewidth]{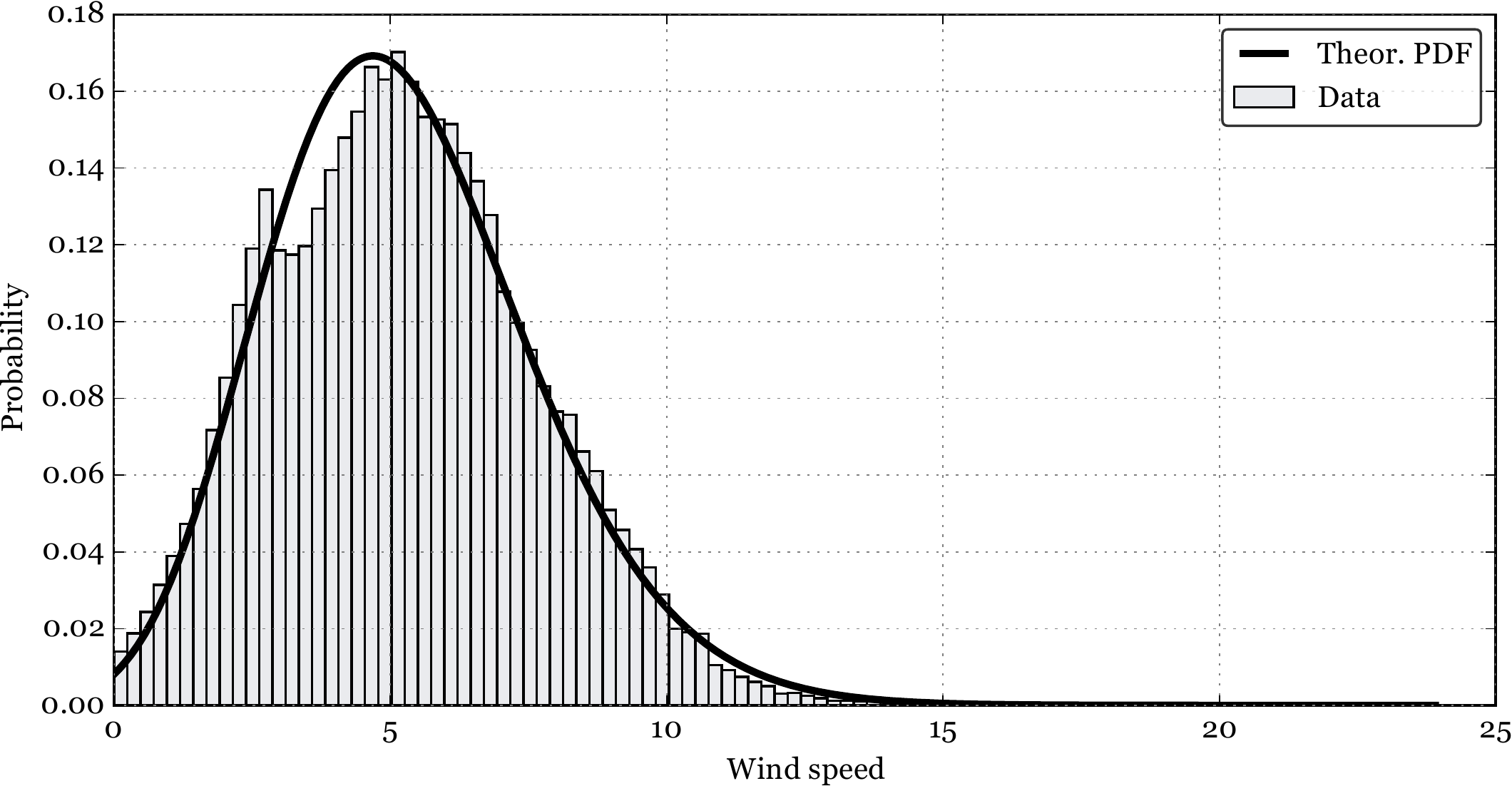}
  \caption{PDF of \textbf{Gamma} distribution compared with data histogram}
  \label{fig:Gamma_PDF}
\end{figure}

\begin{figure}
  \centering
  \includegraphics[width=0.8\linewidth]{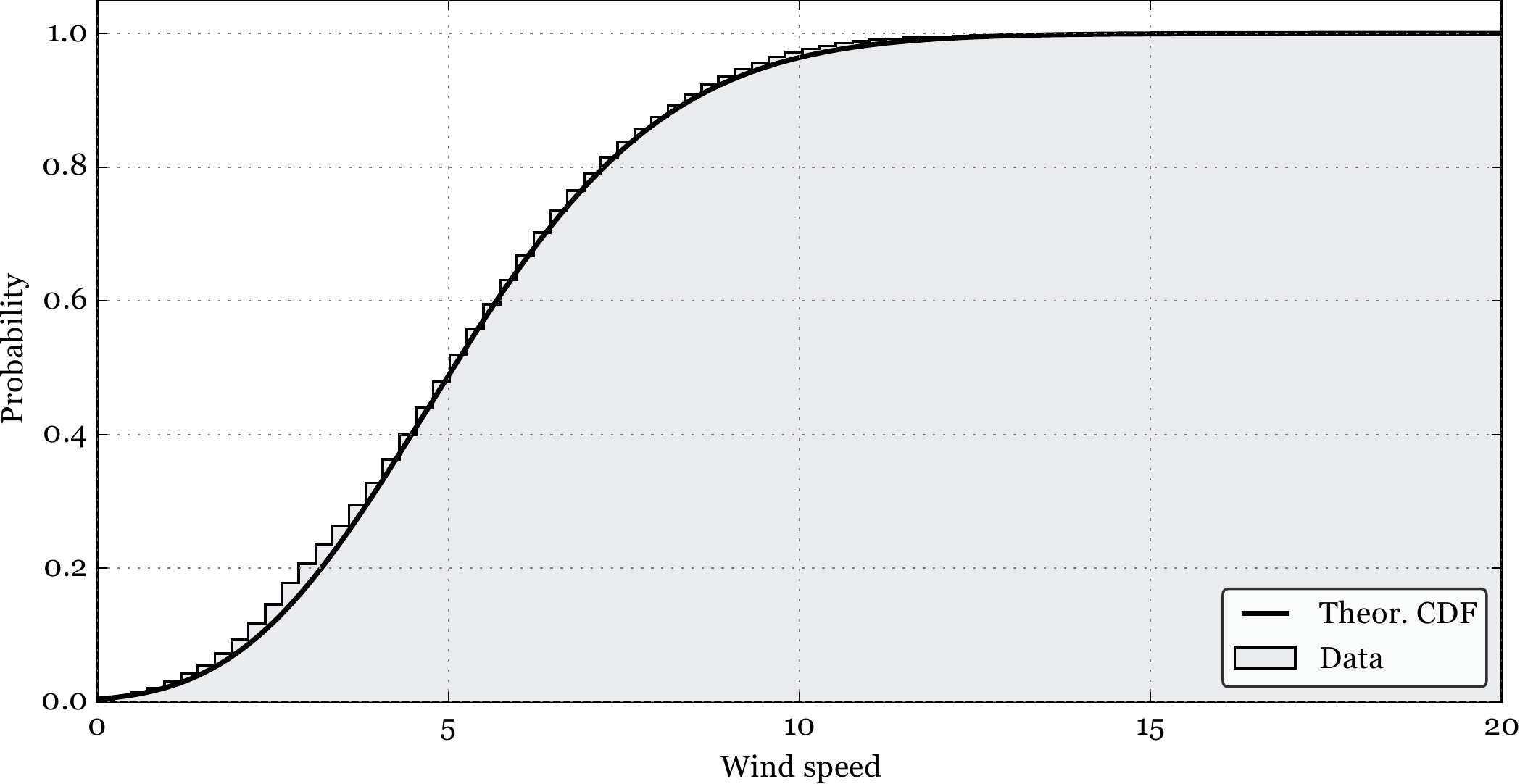}
  \caption{CDF of \textbf{Gamma} distribution compared with empirical distribution function}
  \label{fig:Gamma_CDF}
\end{figure}

\begin{figure}
  \centering
  \includegraphics[width=0.8\linewidth]{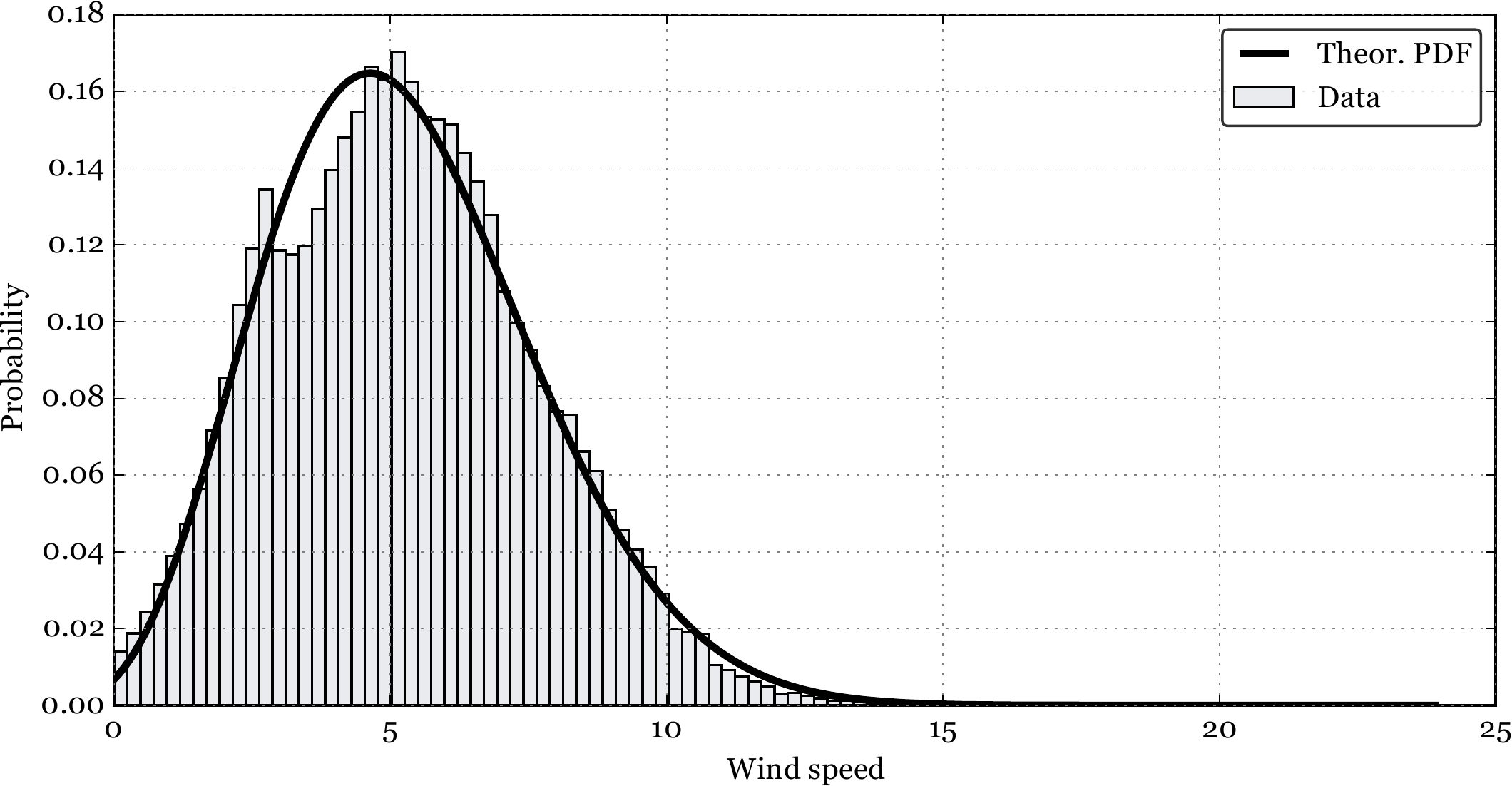}
  \caption{PDF of \textbf{Beta} distribution compared with data histograms}
  \label{fig:Beta_PDF}
\end{figure}

\begin{figure}
  \centering
  \includegraphics[width=0.8\linewidth]{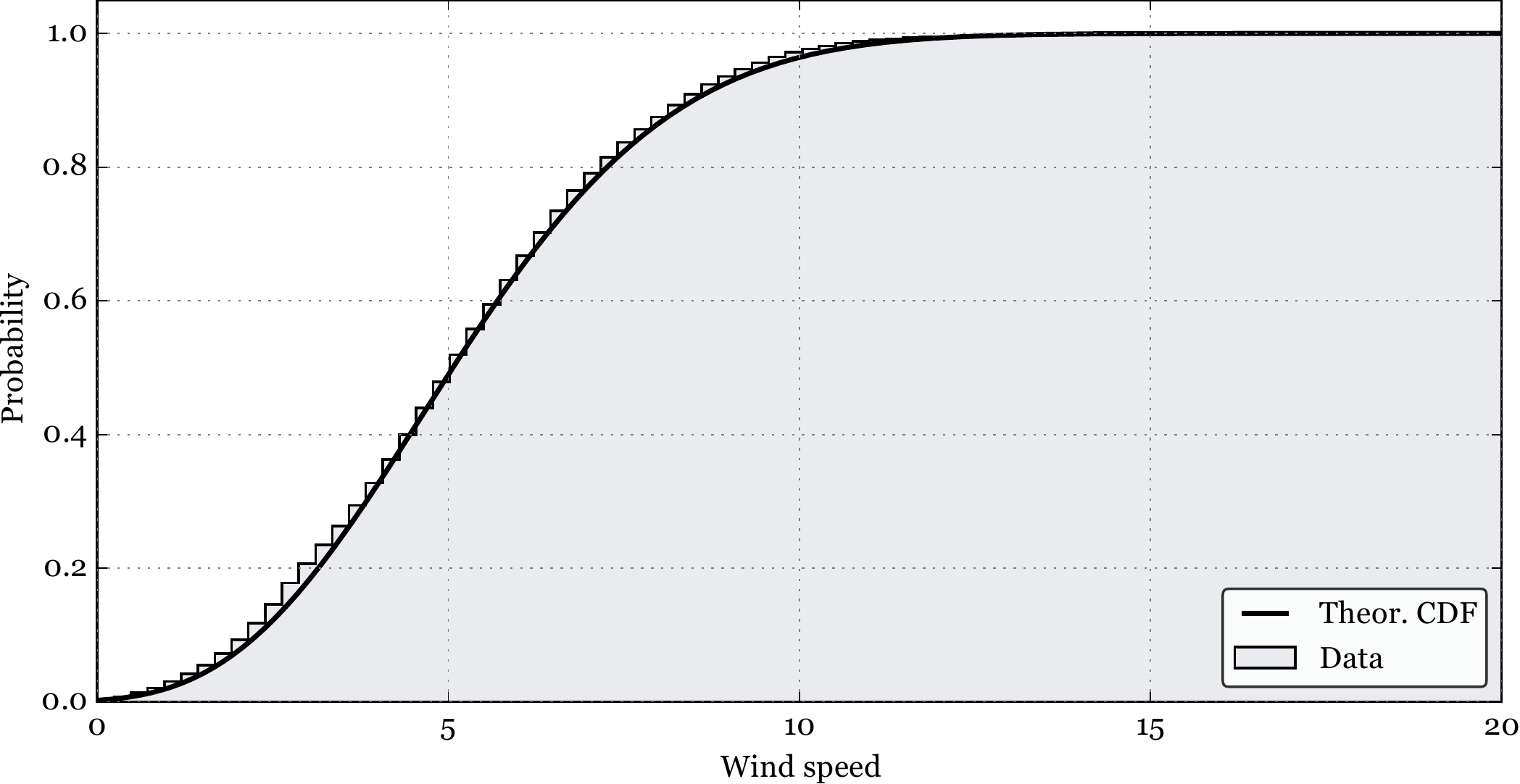}
  \caption{CDF of \textbf{Beta} distribution compared with empirical distribution function}
  \label{fig:Beta_CDF}
\end{figure}

\begin{figure}
  \centering
  \includegraphics[width=0.8\linewidth]{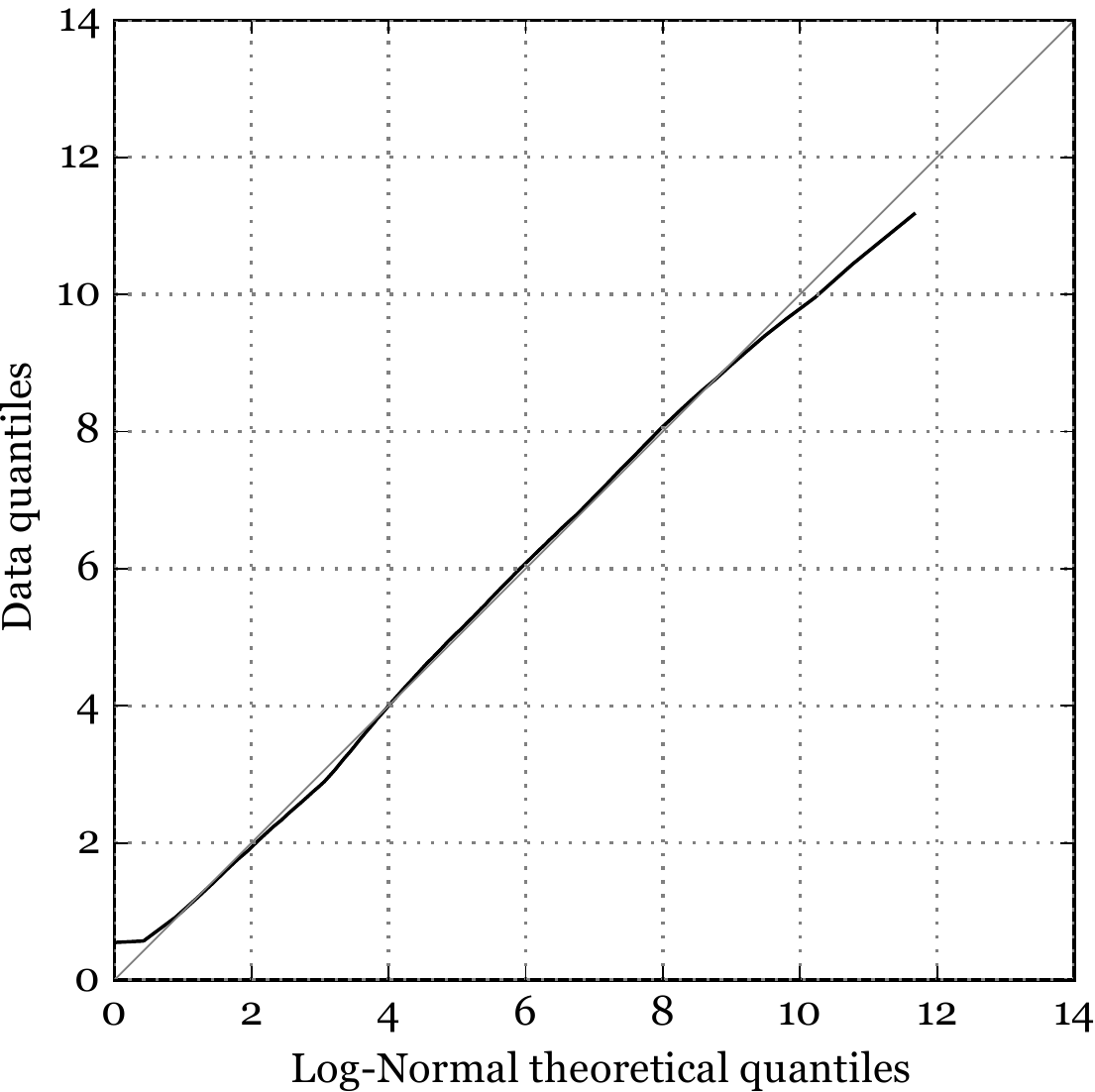}
  \caption{Q-Q plot for LogNormal distribution}
  \label{fig:LogNorm_qqplot}
\end{figure}

\begin{figure}
  \centering
  \includegraphics[width=0.8\linewidth]{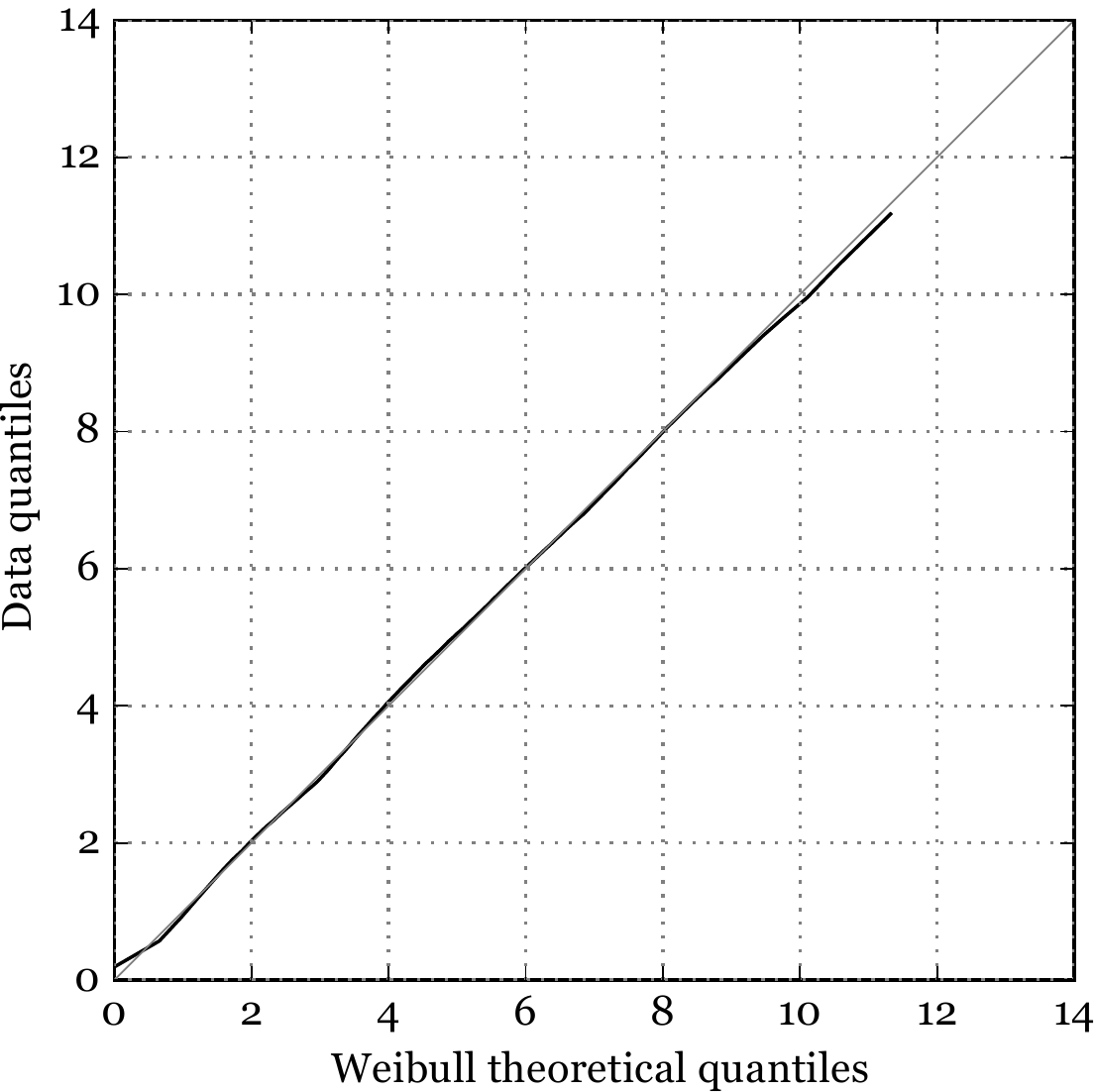}
  \caption{Q-Q plot for Weibull distribution}
  \label{fig:Weibull_qqplot}
\end{figure}

\begin{figure}
  \centering
  \includegraphics[width=0.8\linewidth]{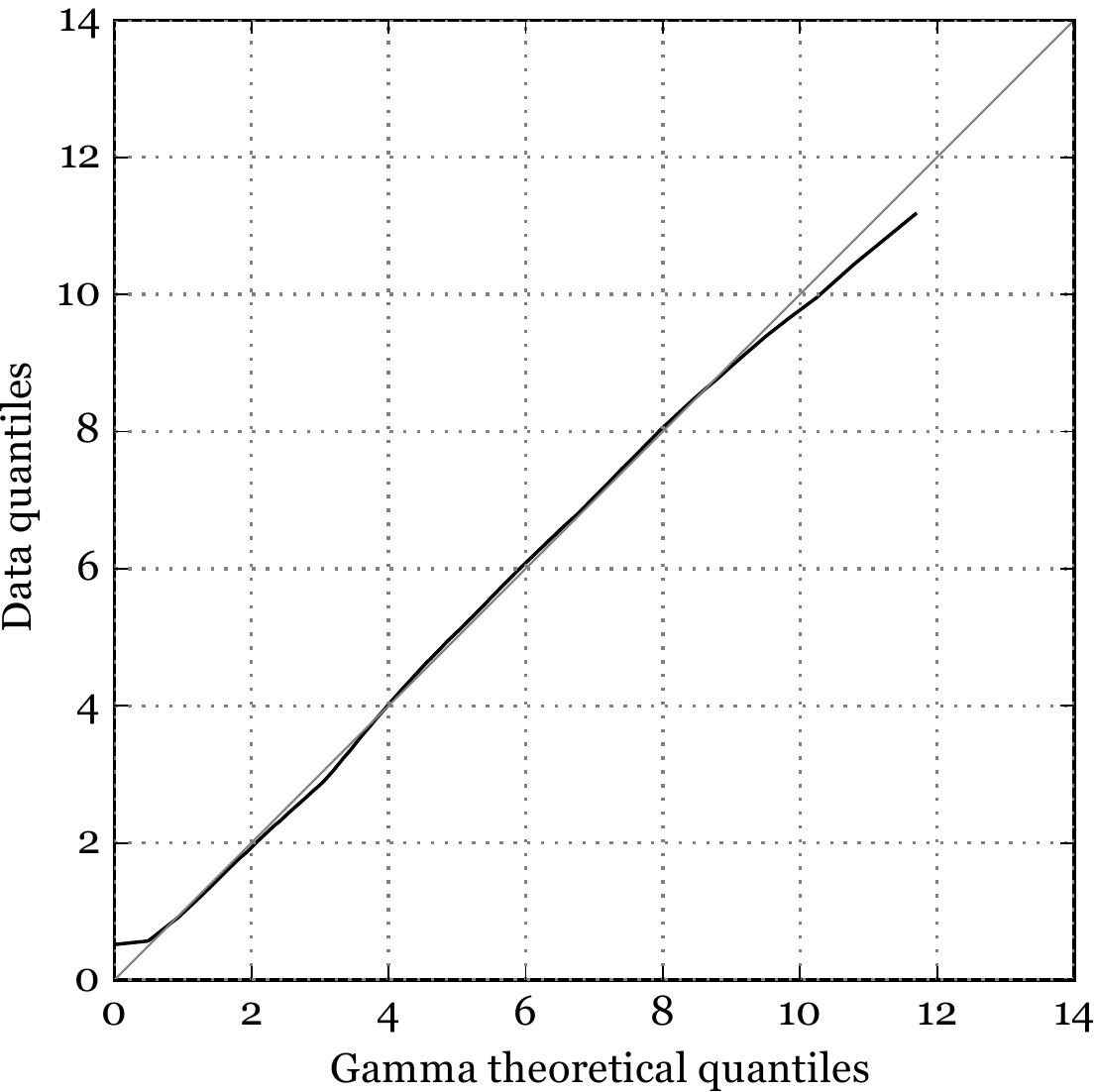}
  \caption{Q-Q plot for Gamma distribution}
  \label{fig:Gamma_qqplot}
\end{figure}

\begin{figure}
  \centering
  \includegraphics[width=0.8\linewidth]{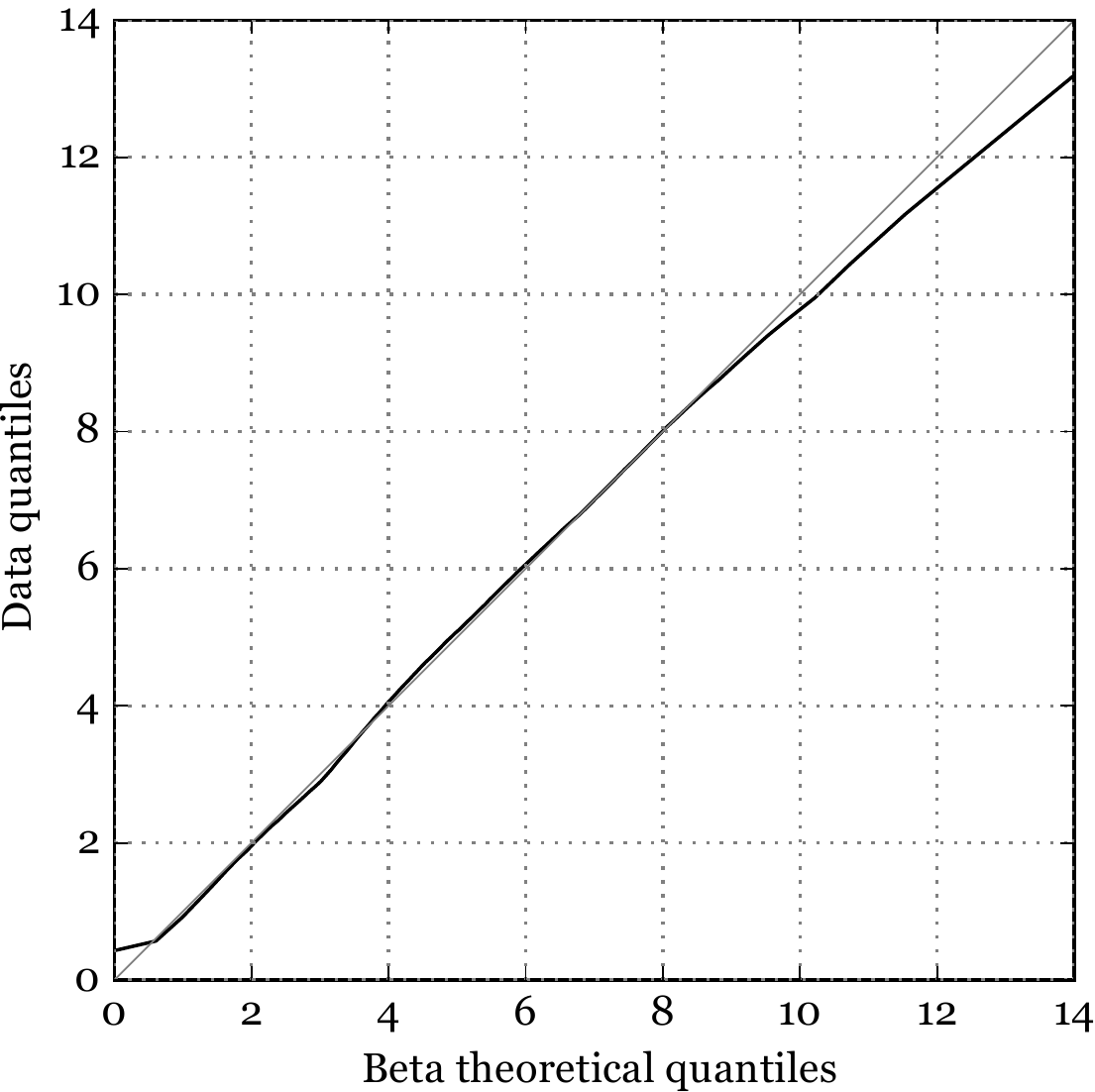}
  \caption{Q-Q plot for Beta distribution}
  \label{fig:Beta_qqplot}
\end{figure}

Результаты исследования представлены в графическом виде на рисунках~\ref{fig:lognormal_PDF},~\ref{fig:lognormal_CDF},
~\ref{fig:Weibull_PDF},~\ref{fig:Weibull_CDF},~\ref{fig:Gamma_PDF},~\ref{fig:Gamma_CDF},~\ref{fig:Beta_PDF},
~\ref{fig:Beta_CDF} и \ref{fig:Weibull_qqplot},~\ref{fig:LogNorm_qqplot},~\ref{fig:Gamma_qqplot},~\ref{fig:Beta_qqplot}. 
Данные изображения были получены для теоретических распределений, параметры которых определялись на основе всего массива данных. 
Из рассмотрения квантиль-квантиль графиков можно сделать вывод, что распределение Вейбулла наилучшим образом подходит 
для аппроксимации имеющихся у нас данных, хотя и превосходит другие распределения лишь незначительно, обходя их лишь 
в точности аппроксимации крайних значений случайной величины.

Мы также провели вычисления с рассматриваемыми распределениями параметризованными только двумя параметрами (положив $l=0$, 
а в бета распределении дополнительно $s=1$). После представления результатов в графическом виде очевидно явное 
превосходство двухпараметрического распределения Вейбулла над остальными двухпараметрическими распределениями 
(логнормального, гамма и бета), чего не наблюдается в трехпараметрическом случае.

\section{Заключение}
Результаты проделанной статистической обработки данных соответствуют результатам, изложенным в литературе, где распределение 
Вейбулла наиболее часто используется для аппроксимации скорости ветра~\cite{L_Lun2000145,L_Seguro200075,L_1983WiEng,L_4488041,L_Islam2011985,L_Garcia1998139}.

Дальнейшая наша работа будет направленна на построение стохастической модели, способной аппроксимировать скорость ветра в 
зависимости от времени~\cite{L_1511.02345}. С другой стороны, мы рассчитываем верифицировать результаты данной работы на 
статистических данных с гораздо большим числом записей, чем имеющиеся в нашем распоряжении на данный момент времени.
\label{sec:conclusion}

\def\acknowledgmentsname{Благодарности}

\begin{acknowledgments}

Работа частично поддержана грантами РФФИ \textnumero~15-07-08795, 16-07-00556.
Также публикация выполнена при финансовой поддержке Минобрнауки России
(Соглашение \textnumero~02.a03.21.0008). 
Расчёты проведены на вычислительном кластере «Felix» РУДН и на
Гетерогенном вычислительном кластере «HybriLIT» Многофункционального
центра хранения, обработки и анализа данных ОИЯИ.

\end{acknowledgments}

  \ifthenelse{\boolean{xetex}}
  {\bibliographystyle{ugost2008l}}
  {\bibliographystyle{gost2008l}}

\bibliography{bib/weibull-test/cite}